\documentclass[a4paper,11pt]{article}
\pdfoutput=1 

\usepackage{jcappub} 

\usepackage[T1]{fontenc} 
\usepackage{eucal}
\usepackage{xspace}
\usepackage{color}
\usepackage{url}
\usepackage{afterpage}

\title{ PyUltraLight: A Pseudo-Spectral Solver for Ultralight Dark Matter Dynamics}

\author[]{Faber Edwards,}
\author[]{Emily Kendall,}
\author[]{Shaun Hotchkiss,}
\author[]{and Richard Easther}

\affiliation[]{Department of Physics, University of Auckland, Private Bag 92019, Auckland, New Zealand}

\emailAdd{faberedwards@gmail.com}
\emailAdd{eken000@aucklanduni.ac.nz}
\emailAdd{s.hotchkiss@auckland.ac.nz}
\emailAdd{r.easther@auckland.ac.nz}

\newcommand{\PyUltraLight}{\textsc{PyUltraLight}\xspace}

\abstract{\PyUltraLight  simulates the dynamics of ultralight dark matter in a non-expanding background. \PyUltraLight can describe the evolution of several interacting ultralight dark matter halos  or one or more halos orbiting a central, fixed Newtonian potential, the latter scenario corresponding to dwarf galaxies orbiting a massive central galaxy. We verify  \PyUltraLight  by showing that it reproduces qualitative dynamical features of previously published simulations and demonstrate that it has  excellent energy-conservation properties.   \PyUltraLight is implemented in a Python-based Jupyter notebook, solving the Schr{\"o}dinger-Poisson equation  governing  ultralight scalar field dark matter dynamics in the non-relativistic regime using a symmetrised split-step pseudospectral  algorithm. The notebook interface makes it simple to specify simulation parameters and  visualise the resulting output but performance-critical routines are managed via calls to computationally efficient compiled libraries. \PyUltraLight  runs on standard desktop hardware with support for shared memory mutlithreading and is available on GitHub.}

\begin{document}
\maketitle
\flushbottom

\section{Introduction}
\label{sec:intro}

The realisation that there may be more to the universe than meets the eye is one of the most profound developments in 20$^{\textrm{th}}$ Century astronomy and astrophysics. However, while there are now multiple lines of evidence that dark matter outweighs baryonic matter by a ratio of approximately 5:1 \cite{Ade:2015xua}, we have few clues regarding the physical nature of dark matter.  Much theoretical and experimental effort has focused on  WIMP models, motivated by their consistency with supersymmetric extensions to the Standard Model and their relatively simple dynamics. However, advanced direct-detection experiments are putting increasingly tight constraints on the WIMP parameter space \cite{Tan:2016zwf, Akerib:2016vxi} and $\Lambda$CDM cosmology with simple, pressureless, noninteracting dark matter (a class including simple WIMP scenarios) is potentially at odds with observations at small astrophysical scales \cite{Bull:2015stt}.  

The potential shortcomings of simple cold dark matter scenarios motivate investigations of more novel dark matter scenarios. In particular, ultralight dark matter (ULDM), also known as fuzzy dark matter (FDM), or BEC dark matter, is an increasingly well-studied possibility; for a recent review of the potential advantages and characteristic attributes of this scenario see Ref~\cite{Hui:2016ltb}. ULDM models are well motivated by fundamental theories possessing approximate shift symmetries such as the theory of the QCD axion \cite{Kim:2008hd, Marsh:2015xka}. Moreover, ULDM can naturally resolve the small-scale problems of $\Lambda$CDM as the Heisenberg uncertainty principle suppresses gravitational collapse on length scales shorter than the de Broglie wavelength of the ULDM particle. In this regime the mass of the ULDM particle becomes correlated with astrophysical observables; if it is on the order of $10^{-22}$~eV, structure is suppressed at kiloparsec scales and below \cite{Hu:2000ke}.

Given the presence of a fundamental  lengthscale, the behaviour of ULDM is more complex than that of simple dark matter scenarios whose cosmologically relevant interactions are purely gravitational. Physically, the effective short-scale pressure and  condensate-like properties of ULDM  create new dynamical possibilities for ULDM scenarios, such as  purely pressure supported soliton-like solutions  \cite{Marsh:2015wka} and superposition or interference during interactions between condensate-like halos \cite{Schwabe:2016rze}. Consequently, modelling dark matter dynamics in ULDM scenarios is more challenging than in simple cold dark matter models, but is critical to understanding the physical consequences of ULDM models. 

In the non-relativistic regime, the dynamics of ULDM can be reduced to the Schr{\"o}dinger-Poisson  system, where the complex variable $\psi$ describes the local density of ULDM quanta while the Poisson equation describes the local gravitational potential. Many approaches have been taken to this problem, including both modifications of existing cosmological simulation codes and the development of new codes specifically designed for ULDM systems. One widely used approach is the Madelung fluid formulation of the Schr{\"o}dinger-Poisson system \cite{Madelung1926} which has a quantum pressure term that can be treated numerically in a variety of ways. In Ref.~\cite{Zhang:2016uiy}, the cosmological code \textsc{gadget} \cite{Springel:2005mi} is modified to treat the quantum pressure as an effective particle-particle interaction and the resulting code, \textsc{axion-gadget} is publicly available \cite{axion-gadget}.  Ref.~\cite{Nori:2018hud} modifies a non-public extension of \textsc{gadget}, \textsc{p-gadget3} to treat the quantum pressure term via smoothed-particle hydrodynamics (SPH) routines. The SPH approach is also  used in Ref.~\cite{Mocz:2015sda}, while a particle-mesh approach was implemented in \cite{Veltmaat:2016rxo}. N\textsc{yx} \cite{Almgren:2013sz} was modified in \cite{Schwabe:2016rze} to study merging ULDM solitonic cores, G\textsc{alacticus} \cite{Benson:2010kx} was modified in \cite{Du:2016zcv} to study the effects of tidal stripping and dynamical friction on ULDM halos, \textsc{arepo} \cite{Springel:2009aa} was modified in \cite{Mocz:2017wlg} to study the core-mass relationship and turbulence characteristics of ULDM halos, and \textsc{gamer} \cite{Schive:2009hw, gamer} was modified \cite{Schive:2014dra} to perform a detailed study of structure formation in ULDM cosmologies. 

While a large number of public codes can solve conventional dark matter scenarios, \textsc{axion-gadget} is the only currently available solver for ULDM dynamics. This paper introduces \PyUltraLight, a stand-alone Python-based pseudospectral Schr{\"o}dinger-Poisson solver, and  demonstrates that it reproduces many of the key findings of more complicated cosmological simulation codes within a desktop computing environment. We anticipate that as a publicly available resource, \PyUltraLight\ will serve as a valuable cross-check on more complex implementations, serve as a basis for further development of such codes within the computational cosmology community, and facilitate explorations of ULDM dynamics.   

\PyUltraLight is based on a symmetrised-split-step (leapfrog) solver for the time evolution, and uses a pseudospectral Fourier algorithm to solve the Poisson equation for the gravitational potential at each step.\footnote{A similar methodology was described in Ref.~\cite{Paredes:2015wga}; at the time of writing this code has not been released. Spectral methods are often used to solve the Poisson equation in large scale structure simulations, while the {\sc PSpectre} code \cite{Easther:2010qz} provides a pseudospectral solver for the evolution of the inflaton  and fields coupled to it during parametric resonance and preheating after inflation \cite{Amin:2010dc,Amin:2011hj,Zhou:2013tsa}.} This algorithm has $2^{nd}$ order accurate time integration steps and sub-percent level energy conservation, while the wavefunction normalisation  is conserved to machine precision. As a pseudospectral code, linear differential operators are computed by direct multiplication in the Fourier domain, while non-linear terms are evaluated in position space. Consequently, \PyUltraLight is free from noise associated with spatial derivatives computed via finite-differencing. There is a necessary computational cost associated with the Fourier and inverse Fourier transforms but these transforms are optimised in \PyUltraLight through the use of the pyFFTW pythonic wrapper around the C-based FFTW subroutine library \cite{pyfftw,fftw}. As the FFTW libraries offer full parallelisation, \PyUltraLight is currently designed to take advantage of multiple cores on a user PC or shared-memory environment. Full MPI compatibility has not yet been implemented as we have not found a need to run simulations in a distributed-memory cluster environment, however future releases may address this possibility. 

This paper is organised as follows. We first provide a short review of ULDM physics, including a derivation of the Schr{\"o}dinger-Poisson equations from the underlying scalar-field Lagrangian. We then describe their implementation in \PyUltraLight, before moving on to describe testing and verification procedures applied to the code. We reproduce a selection of results from a variety of recent ULDM simulations and discuss the energy conservation and accuracy as a function of spatial resolution.

\section{The Physics of ULDM}\label{sec:physics}

\subsection{The Schr{\"o}dinger-Poisson System}

The existence of an extremely light scalar field, minimally coupled to gravity, is the central premise on which ULDM models are predicated. Within the ULDM framework, it is proposed that this scalar field exists as a Bose-Einstein condensate, described by a single wavefunction which is governed by the Schr{\"o}dinger-Poisson coupled differential equations. We begin by deriving this system of equations as a non-relativistic weak-field limit of a more general theory. We start from the action functional for a scalar field, $\phi$, minimally coupled to gravity and in the absence of self-interactions,
\begin{equation}\label{eq:action}
    S=\int \frac{d^4x}{\hbar}\sqrt{-g}\bigg\{\frac{1}{2}g^{\mu\nu}\partial_\mu\phi\partial_\nu\phi-\frac{m^2}{2\hbar^2}\phi^2\bigg\},
\end{equation}
where we have taken $c=1$ but retain factors of $\hbar$ at this stage. Applying the variational principle to this action yields the Euler-Lagrange equations
\begin{equation}\label{eq:e-l}
    \frac{1}{\sqrt{-g}}\partial_\mu\big[\sqrt{-g}g^{\mu\nu}\partial_\nu\phi\big]-\frac{m^2}{\hbar^2}\phi=0.
\end{equation}
We evaluate equation \ref{eq:e-l} using linear perturbation theory, adopting the perturbed FRW metric in the Newtonian gauge:
\begin{equation}\label{eq:pFRW}
    ds^2=-\big(1+2\Phi(\vec{r},t)\big)dt^2+a(t)^2\big(1-2\Phi(\vec{r},t)\big)d\vec{r}^{\,2}.
\end{equation}
To linear order in $\Phi(\vec{r})$ we obtain 
\begin{equation}\label{eq:linear}
    \Ddot{\phi}-\frac{\big(1+4\Phi\big)}{a(t)^2}\nabla^2\phi+3H\Dot{\phi}-4\dot{\Phi}\dot{\phi}+\big(1+2\Phi\big)\frac{m^2}{\hbar^2}\phi=0,
\end{equation}
where $H=\Dot{a}(t)/a(t)$. At late times in an expanding universe, $H\ll m/\hbar$ and it is sufficient to set $H=0$ and $a(t)=1$ in equation \ref{eq:linear}. This is a good approximation even at relatively high redshifts, including the epochs of early structure formation. Alternatively, if we consider a non-expanding universe, these equalities are true by definition. In either case, we can remove the third term in equation \ref{eq:linear}. The resulting equation can then be analysed using WKB methods in the non-relativistic regime.\footnote{For a detailed explication of the WKB approximation in the non-relativistic limit, see \cite{Young2015}.} This allows us to write an ansatz solution for the field $\phi$:
\begin{equation}\label{eq:ansatz}
    \phi=\frac{\hbar}{\sqrt{2}m}\big(\psi e^{-imt/\hbar}+\psi^* e^{imt/\hbar}\big),
\end{equation}
where $\psi$ is assumed to be slowly varying in the sense that $m\vert\psi\vert\gg\hbar\vert\dot{\psi}\vert$, $m\vert\dot{\psi}\vert\gg\hbar\vert\Ddot{\psi}\vert$, $m\vert\psi\vert\gg\hbar\vert\nabla{\psi}\vert$, and $m\vert\nabla{\psi}\vert\gg\hbar\vert\nabla^2{\psi}\vert$. Since $\Phi$ is sourced by $\psi$, we also have that $m\vert\Phi\vert\gg\hbar\vert\dot{\Phi}\vert$. Direct substitution of the ansatz solution into equation \ref{eq:linear}, discarding heavily suppressed terms, yields
\begin{equation}\label{eq:schrodinger}
    i\hbar\Dot{\psi}=-\frac{\hbar^2}{2m}\nabla^2\psi+m\Phi\psi.
\end{equation}
We have thus shown that $\psi$ satisfies the Schr{\"o}dinger equation in this limit, which is interpreted as the macroscopic wavefunction of a Bose-Einstein condensate. It follows that the particle number density of the condensate is given by $\vert\psi\vert^2$, so its mass density is simply $m\vert\psi\vert^2$. The local gravitational potential thus satisfies the Poisson equation,
\begin{equation}\label{eq:poisson}
    \nabla^2\Phi=4\pi G m \vert\psi\vert^2,
\end{equation}
where $G$ is Newton's gravitational constant. The coupled equations \ref{eq:schrodinger} and \ref{eq:poisson} together form the nonlinear Schr{\"o}dinger-Poisson system which describes the dynamics of ULDM in the non-relativistic regime. While Equations \ref{eq:schrodinger} and \ref{eq:poisson} are valid for open boundary conditions, \PyUltraLight is designed to solve the Schr{\"o}dinger-Poisson system under periodic boundary conditions. In this case the correct form of equation \ref{eq:poisson} is 
\begin{equation}\label{eq:poisson_periodic}
    \nabla^2\Phi=4\pi G m \big(\vert\psi\vert^2-\langle\vert\psi\vert^2\rangle\big),
\end{equation}
where we subtract the average density from the right hand side of the Poisson equation. The form of Equation \ref{eq:poisson_periodic} is a consequence of Gauss' law and the fact that the surface integral of the gradient of the field around the perimeter of the simulation grid is identically zero when periodic boundary conditions are imposed \cite{Dabo:2008}.

\subsection{Field Rescalings}

It is helpful to recast the Schr{\"o}dinger-Poisson system (equations \ref{eq:schrodinger} and \ref{eq:poisson}) in terms of adimensional quantities. In keeping with Refs~\cite{Schive:2014hza,Paredes:2015wga} we introduce length, time, and mass scales as follows:
\begin{align}
    \CMcal{L}&=\left(\frac{8\pi\hbar^2}{3 m^2H_0^2\Omega_{m_0}}\right)^{\frac{1}{4}}\approx121\left(\frac{10^{-23}\operatorname{eV}}{m}\right)^{\frac{1}{2}}\operatorname{kpc},\label{eq:length}\\
    \CMcal{T}&=\left(\frac{8\pi}{3 H_0^2\Omega_{m_0}}\right)^{\frac{1}{2}}\approx75.5 \operatorname{Gyr},\label{eq:time}\\
    \CMcal{M}&=\frac{1}{G}\left(\frac{8\pi}{3 H_0^2\Omega_{m_0}}\right)^{-\frac{1}{4}}\left(\frac{\hbar}{m}\right)^{\frac{3}{2}}\approx 7\times 10^7\left(\frac{10^{-23}\operatorname{eV}}{m}\right)^{\frac{3}{2}}\operatorname{M}_{\odot},\label{eq:mass}
\end{align}
where $m$ is the mass of the ultralight scalar field, $H_0$ is the present-day Hubble parameter, $G$ is Newton's gravitational constant and $\Omega_{m_0}$ is the present-day matter fraction of the energy density of the universe. We  recast equations \ref{eq:schrodinger} and \ref{eq:poisson} in terms of the  dimensionless quantities
\begin{equation}\label{eq:dimensionless}
    t'=\frac{t}{\CMcal{T}},\quad
    \vec{x}^{\,'}=\frac{\vec{x}}{\CMcal{L}},\quad
    \Phi'=\frac{m\CMcal{T}}{\hbar}\Phi,\quad
    \psi'=\CMcal{T}\sqrt{mG}\psi.
\end{equation}
Dropping the primes for notational convenience, we see that the coupled differential equations of the  Schr{\"o}dinger-Poisson system under periodic boundary conditions reduce to 
\begin{align}
    i\Dot{\psi}(\vec{x},t)&=-\frac{1}{2}\nabla^2\psi(\vec{x},t)+\Phi(\vec{x},t)\psi(\vec{x},t),\label{eq:s-adim}\\
    \nabla^2\Phi(\vec{x},t)&=4\pi\big(\vert\psi(\vec{x},t)\vert^2-\langle\vert\psi(\vec{x},t)\vert^2\rangle\big),\label{eq:p-adim}
\end{align}
where it is understood that all quantities involved are dimensionless. We can recover  dimensionful quantities via the ``dictionary'' provided by equations \ref{eq:length} to \ref{eq:mass}. For example, the integrated mass of the system, $M_{tot}$, is given by
\begin{equation}\label{eq:integrated-mass}
    M_{tot}=\CMcal{M}\int d^3x\vert\psi\vert^2.
\end{equation}
Likewise, the mass density at any point is given by
\begin{equation}\label{eq:density}
    \rho=\CMcal{M}\CMcal{L}^{-3}\vert\psi\vert^2.
\end{equation}
By dimensional analysis, we can easily restore dimensionful units to any of the quantities calculated by the code. In particular, in the following sections it is to be understood that
\begin{equation}
    E=\CMcal{M}\CMcal{L}^2\CMcal{T}^{-2}\ E_{code}, \quad v=\CMcal{L}\CMcal{T}^{-1}\ v_{code}, 
\end{equation}
where $E$ and $v$ represent energy and velocity, respectively. \PyUltraLight\ works internally with these dimensionless quantities but can receive initial conditions and generates output in physical units. Henceforth, we will often refer to $\vert\psi\vert^2$  as the density, where it is understood that this is in fact a dimensionless quantity related to the physical mass density via the constant of proportionality given by equation \ref{eq:density}.

\section{Algorithm and Implementation}\label{sec:implementation}

In this section we discuss the methodology used to calculate the dynamics of the adimensional  Schr{\"o}dinger-Poisson system (equations \ref{eq:s-adim} and \ref{eq:p-adim}) given user-specified initial conditions. We introduce the symmetrised split-step Fourier method, and schematically describe  how the system is evolved at each timestep. 

\subsection{Dynamical Evolution}\label{sec:dynamics}

Dynamical evolution within \PyUltraLight\ progresses via a symmetrised split-step Fourier process on an $N\times N\times N$  grid with  periodic spatial boundary conditions. To understand this method,  first consider the exact expression for the unitary time evolution of the wavefunction according to equation \ref{eq:s-adim}, namely
\begin{equation}\label{eq:exact-time-ev}
    \psi(\vec{x},t+h)=\mathcal{T}\operatorname{exp}\left[-i\int_t^{t+h}dt'\left\{-\frac{1}{2}\nabla^2+\Phi(\vec{x},t')\right\}\right]\psi(\vec{x},t),
\end{equation}
where $\mathcal{T}$ is the time-ordering symbol. For a sufficiently small timestep $h$,  the trapezoidal rule gives the approximation
\begin{equation}
    \int_t^{t+h}dt'\Phi(\vec{x},t')\approx \frac{h}{2}\Big(\Phi(\vec{x},t\hspace*{-.2em}+\hspace*{-.2em}h)+\Phi(\vec{x},t)\Big).
\end{equation}
We can therefore write the approximate form of equation \ref{eq:exact-time-ev} as
\begin{equation}\label{eq:approx-time-ev}
    \psi(\vec{x},t+h)\approx \operatorname{exp}\left[i\frac{h}{2}\Big(\nabla^2-\Phi(\vec{x},t\hspace*{-.2em}+\hspace*{-.2em}h)-\Phi(\vec{x},t)\Big)\right]\psi(\vec{x},t).
\end{equation}
Note that the exponential in equation \ref{eq:approx-time-ev} omits the time-ordering symbol, and is only equivalent to its time-ordered counterpart to order $h^2$. 

The linear differential operator in equation \ref{eq:approx-time-ev} acts naturally in Fourier space, while the nonlinear potential term is simplest to evaluate in position space. By splitting the exponential we can evaluate each term in its natural domain. Such a splitting is valid when the timestep is small, and is represented as
\begin{equation}\label{eq:split-step}
    \psi(\vec{x},t\hspace*{-.2em}+\hspace*{-.2em}h)\approx\operatorname{exp}\left[-\frac{ih}{2}\Phi(\vec{x},t\hspace*{-.2em}+\hspace*{-.2em}h)\right]\operatorname{exp}\left[\frac{ih}{2}\nabla^2\right]\operatorname{exp}\left[-\frac{ih}{2}\Phi(\vec{x},t)\right]\psi(\vec{x},t).
\end{equation}
This splitting can be understood thusly: first, a half timestep is taken in which only the nonlinear potential operator acts, followed by a full timestep in the linear term. The potential field is then updated, and a final half timestep in the nonlinear term is performed. Using the Baker-Campbell-Hausdorff formula to express the product of exponentials in equation \ref{eq:split-step} as a single exponential and keeping only terms to order $h^2$ we find:
\begin{equation}\label{eq:BCH}
    \operatorname{exp}\left[i\frac{h}{2}\Big(\nabla^2-\Phi(\vec{x},t\hspace*{-.2em}+\hspace*{-.2em}h)-\Phi(\vec{x},t)\Big)+\frac{h^2}{8}\Big[\nabla^2,\Phi(\vec{x},t)\Big]-\frac{h^2}{8}\Big[\nabla^2,\Phi(\vec{x},t\hspace*{-.2em}+\hspace*{-.2em}h)\Big]\right].
\end{equation}
Making use of the fact that $\Phi(\vec{x},t+h)\approx\Phi(\vec{x},t)+h\Dot{\Phi}(\vec{x},t)$ we see that the commutators in equation \ref{eq:BCH} cancel at $\mathcal{O}(h^2)$ and the expression matches \ref{eq:approx-time-ev}, with the dominant error term appearing at $\mathcal{O}(h^3)$. 

Evaluation of equation \ref{eq:split-step} within \PyUltraLight thus proceeds as follows:
Initially, the nonlinear term acts in position space for one half-timestep. The result is  Fourier transformed, and a full timestep is taken with the differential operator  applied in the Fourier domain. The potential field is then updated in accordance with equation \ref{eq:p-adim}. After an inverse Fourier transform a final half timestep is taken with the updated nonlinear term acting in position space to give the new $\psi$ field configuration. This procedure is known as the symmetrised split-step Fourier method, and  used widely in fields such as nonlinear fiber optics \cite{Agrawal2013, Zhang:2008, Sinkin2003}. 

The algorithm can be represented schematically as
\begin{equation}\label{eq:schematic-psi}
    \psi(\vec{x},t\hspace*{-.2em}+\hspace*{-.2em}h)=\operatorname{exp}\left[-\frac{ih}{2}\Phi(\vec{x},t\hspace*{-.2em}+\hspace*{-.2em}h)\right]\CMcal{F}^{-1} \operatorname{exp}\left[\frac{-ih}{2}k^2\right]\CMcal{F}\operatorname{exp}\left[-\frac{ih}{2}\Phi(\vec{x},t)\right]\psi(\vec{x},t),
\end{equation}
where the order of operations runs from right to left, $\CMcal{F}$ and $\CMcal{F}^{-1}$ denote the discrete Fourier transform and its inverse, and $k$ is the wavenumber in the Fourier domain. The potential field is updated following the inverse Fourier transform in equation \ref{eq:schematic-psi}, via
\begin{equation}\label{eq:schematic-phi}
    \Phi(\vec{x},t\hspace*{-.2em}+\hspace*{-.2em}h)=\CMcal{F}^{-1}\left(-\frac{1}{k^2}\right)\CMcal{F}\ 4\pi\vert\psi(\vec{x},t_{i})\vert^2,
\end{equation}
where $\psi(\vec{x},t_{i})$ is the field configuration at this halfway point in the full timestep. We explicitly set the $k=0$ Fourier mode to zero prior to the final inverse Fourier transform; as a consequence there is no need to subtract the global average density from the local value in Equation \ref{eq:schematic-phi}, in contrast to Equation \ref{eq:poisson_periodic}. The final operation in equation \ref{eq:schematic-psi}  only changes the phase of  $\psi$, so we could replace $\psi(\vec{x},t_{i})$ with $\psi(\vec{x},t\hspace*{-.2em}+\hspace*{-.2em}h)$ in equation \ref{eq:schematic-phi} with no change in meaning. \PyUltraLight makes an additional simplification to the symmetrised split-step Fourier method by combining the consecutive half-steps in the nonlinear term into a single full step. Consequently, only the first and last operations involve actual half steps. Schematically this becomes 
\begin{equation}
    \psi(t\hspace*{-.2em}+\hspace*{-.2em}nh)=\operatorname{exp}\left[+\frac{ih}{2}\Phi\right]\left(\prod^n\operatorname{exp}\left[-ih\Phi\right]\operatorname{exp}\left[\frac{ih}{2}\nabla^2\right]\right)\operatorname{exp}\left[-\frac{ih}{2}\Phi\right]\psi(t),
\end{equation}
where  $\Phi$  is updated at each step via equation \ref{eq:schematic-phi}; attention is drawn to the sign difference between the first and last operators.

From a computational perspective, the numerical Fourier transforms are likely to be the rate-limiting step in any pseudospectral code. In \PyUltraLight the discrete Fourier transform (DFT) and its inverse are implemented via pyFFTW, a pythonic wrapper for the C-based FFTW subroutine library which efficiently implements both real and complex DFTs \cite{pyfftw,fftw,Frigo2005}. This  allows  \PyUltraLight  to combine the flexibility of a notebook based modelling tool with the efficiency of a carefully tuned, compiled numerical library.  FFTW is fully parallelised and its support for multithreading  is inherited by pyFFTW and accessed within \PyUltraLight; the number of threads used by the pyfftw.FFTW class is determined by the  Python multiprocessing routines which are used to ascertain the number of available CPU cores. In addition, \PyUltraLight uses the NumExpr package to parallelise operations on array objects within the simulation \cite{numexpr}. 

\subsection{Initial Conditions: Soliton Profiles}\label{sec:soliton-profiles}

\PyUltraLight specifies the initial dark matter configuration as a superposition of an arbitrary number of solitonic halos, with arbitrary (user-defined) velocities and phases. This is necessarily an idealisation, given that realistic dark matter halos will not map directly to the solitonic solutions, but it provides an excellent ``playground'' in which to explore ULDM dynamics, and the initialisation routines within \PyUltraLight can be easily augmented to accommodate a wider range of scenarios. The initial field configuration is built by loading a NumPy array file encoding a solitonic solution to the Schr{\"o}dinger-Poisson system and the corresponding position mass, velocity, and phase parameters each specified by the user within the accompanying Jupyter notebook. 

In practice, only a finite  range of halo masses can be supported within a given simulation -- the  radius of a solitonic halo is inversely proportional to its mass, so resolving a light halo interacting with a very massive halo would require an extremely fine spatial mesh.  However, \PyUltraLight also allows the user to specify a fixed, external potential which does not take part in the dynamics. At this point only a central $1/r$ potential is supported but this would be easily generalised. It should be noted that because \PyUltraLight enforces periodic boundary conditions, care must be taken in cases where solitons approach the boundaries of the simulation grid. If a soliton were to cross the boundary during a simulation in which a Newtonian central potential is implemented, the forces exerted during the crossing would be unphysical. For studies of orbital stability this is unlikely to cause any problems, as in these circumstances material collapses toward the centre of the simulation grid rather than crossing the boundaries. However, the user should ensure that solitons are initialised sufficiently far from the boundary for the purposes of each simulation on a case-by-case basis. In situations where a significant portion of the total mass is expected to be ejected, such as the merger of multiple solitons to form a larger halo, care should be taken to ensure that mass ejected above the escape velocity is not recaptured as it re-enters the grid from the other side. For studies of this kind, an absorbing sponge at the grid boundaries is perhaps more suitable than periodic boundary conditions, though this has not been implemented in \PyUltraLight at this stage. 
 
The soliton profile used to generate the initial conditions in \PyUltraLight is found by first imposing spherical symmetry in the Schr{\"o}dinger-Poisson equations and assuming time independence in the radial density profile \cite{Paredes:2015wga}:
\begin{equation}
    \psi(\vec{x},t)\rightarrow e^{i\beta t} f(r), \quad \Phi(\vec{x},t)\rightarrow \varphi(r),
\end{equation}
where $r=\vert\vec{x}\vert$. Introducing $\tilde{\varphi}(r):=\varphi(r)+\beta$, equations \ref{eq:s-adim} and \ref{eq:p-adim}  reduce to
\begin{align}
    0 &= -\frac{1}{2}f^{\prime\prime}(r)-\frac{1}{r}f^\prime(r)+\tilde{\varphi}(r)f(r)\label{eq:s-spherical}\\
    0 &= \tilde{\varphi}^{\prime\prime}(r)+\frac{2}{r}\tilde{\varphi}^\prime(r)-4\pi f(r)^2\label{eq:p-spherical}
\end{align}
where primes denote derivatives with respect to $r$. Note that this system contains no arbitrary constants, so the underlying profile is effectively universal and is loaded as a pre-computed array by \PyUltraLight, rather than computed from scratch with each code execution. The soliton profile numpy array file is included with \PyUltraLight, however, an auxiliary program {\sc soliton\texttt{\_}solution.py} is also supplied, from which this array can be generated; it uses a  fourth-order Runge-Kutta algorithm to solve the coupled profile equations. We set $\left. f(r) \right|_{r=0}$=1, while  smoothness requires that first derivatives of $f(r)$ and $\tilde{\varphi}(r)$ vanish at the origin. We then use the shooting method to search for solutions of $f(r)$ and $\varphi(r)$  satisfying the boundary conditions $\lim_{r\rightarrow\infty}\varphi(r)=0$ and $\lim_{r\rightarrow\infty}f(r)=0$, varying $\left. \tilde{\varphi}(r)\right|_{r=0}$ until we obtain a solution of $f(r)$ which approaches zero at the maximal specified radius, $r_m$. The value of $\beta$ is then calculated by assuming that $\varphi(r)$ goes as $-c/r$ at large radii, where $c$ is a constant. Under this assumption, we can write
\begin{equation}
    \tilde{\varphi}(r_m)=-\frac{c}{r_m}+\beta, \quad c=r_m^2\tilde{\varphi}^{\ \prime}(r_m).
\end{equation}
We thus obtain the full solution $\psi(\vec{x},t)=e^{i\beta t}f(r)$.
Having initially chosen $\left. f(r)\right|_{r=0}=1$, we may then generalise to $\left. f(r)\right|_{r=0}=\alpha$, where $\alpha$ is an arbitrary positive real number. It is easily verified that if $e^{i\beta t}f(r)$ is a solution to the spherically symmetric Schr{\"o}dinger-Poisson system, then $g(r)$ is also a solution, where
\begin{equation}
    g(r)=e^{i\alpha\beta t}\alpha f(\sqrt{\alpha}\ r).
\end{equation}
We thus have a family of spherically symmetric soliton solutions to the dimensionless Schr{\"o}\-dinger-Poisson system; the dimensionless soliton mass is proportional to $\sqrt{\alpha}$ and the full width at half maximum is proportional to $1/\sqrt{\alpha}$. Since the size of the soliton scales inversely with the mass, the most massive soliton in the solution  puts a lower bound on the required spatial resolution. 

The Schr{\"o}dinger equation is not trivially form invariant under Galilean boosts but we can enforce Galilean covariance through the addition of a velocity-dependent phase factor,
\begin{equation}\label{eq:covariant-solution}
    \psi(\vec{x},t)=\alpha f\big(\sqrt{\alpha}\vert\vec{x}-\vec{v}t\vert\big)e^{i\left(\alpha\beta t+\vec{v}\cdot\vec{x}-\frac{1}{2}\vert\vec{v}\vert^2t\right)}.
\end{equation}
To construct the initial field configuration,  \PyUltraLight, loads the NumPy array encoding the radial profile $f(r)$ for the $\left. f(r)\right|_{r=0}=1$ case. Equation \ref{eq:covariant-solution} is then used to transform the this solution into soliton(s) with user-specified values position, mass, and velocity specified, via the accompanying Jupyter notebook. The user may also add an additional constant phase factor if desired.

\subsection{Choosing the Timestep}

The Courant-Friedrichs-Lewy (CFL) condition is an upper bound on the timestep (as a function of grid-spacing) that must be satisfied by many partial differential equation solvers based on finite-differencing  \cite{Ajaib:2013oua} and is often cited in numerical analyses of ULDM via the Schr{\"o}dinger-Poisson system, see e.g. Ref.~\cite{Schwabe:2016rze}. However, the CFL condition expresses a causality constraint, and is generally only strictly applicable to hyperbolic PDEs, whereas the Schr{\"o}dinger-Poisson has only a first order time derivative, even though it is effectively the nonrelativistic limit of the Klein-Gordon equation. Moreover, because \PyUltraLight computes spatial derivatives via a pseudospectral method, technically it is unconditionally stable \cite{Taha:1984jz}. Our split-step algorithm is  second order in the timestep, and its value will always be an empirical tradeoff between computational cost and convergence to the apparent limit in which step is arbitrarily small. Consequently, the user is encouraged to validate their choice  of timestep via case-by-case  convergence testing. 

The default timestep in  \PyUltraLight  is fixed with reference to the fluid interpretation of the Schr{\"o}dinger-Poisson system \cite{Hui:2016ltb}. The fluid interpretation is often used to recast the Schr{\"o}dinger-Poisson system in the form of the Madelung equations \cite{Suarez:2011yf}, which a hydrodynamical representation of the system. The first step is to define
\begin{equation}\label{eq:madelung}
    \psi\equiv\sqrt{\rho}e^{i\theta}, \quad \vec{v}\equiv\boldsymbol{\nabla}\theta.
\end{equation}
and to treat $\vec{v}$ as a fluid velocity.  From this perspective, if the phase difference between two adjacent grid points exceeds $\pi$ the fluid will appear to move ``backwards'' across the grid. We thus set the default timestep, $\Delta t$, so that fluid travelling at this maximum velocity traverses one grid space, $\Delta x$, per timestep, or
\begin{equation}\label{eq:timestep}
    \Delta t = \frac{(\Delta x)^2}{\pi}.
\end{equation}
This is a choice, rather than a strict constraint on $\Delta t$. However, if the ``fluid'' approaches velocities where the phase appears to switch direction, the configuration is approaching the point where the simulation grid is too coarse to fully resolve the dynamics. Hence, a timestep much smaller than this value may offer little practical advantage. However, in some cases the breakdown may occur in regions of the simulation volume that are of little physical interest, and the user is free to choose a larger timestep via the 'step\textunderscore factor' parameter in the Jupyter notebook.

Alternatively, Ref.~\cite{Mocz:2017wlg} fixes the timestep by ensuring that neither of the unitary operators in Equation \ref{eq:schematic-psi}  lead to a phase change of more than $\pi$ for a single grid point over one timestep. However, because the pseudo-spectral algorithm does not compare the phase of a single gridpoint at different points in time, this choice of timestep is not a requirement for stability. This method gives the following constraints: 
\begin{equation}
    \Delta t < \left[\frac{2\pi}{\Phi_{max}}, \frac{2(\Delta x)^2}{\pi}\right],
\end{equation}
where the second of these constraints is generally the stricter of the two, and is equivalent to our default choice of timestep up to a multiplicative factor of $\mathcal{O}(1)$. Our experience is that specifying the timestep via Equation \ref{eq:timestep} is suitable for the majority of simulation scenarios, and we explore convergence in more detail in Section \ref{sec:resolution}.

\section{ULDM Dynamics with \PyUltraLight}\label{sec:test}

In this section we validate \PyUltraLight by reproducing results from previous studies of ULDM dynamics, demonstrating interference effects and effective repulsive forces arising from the wavelike nature of ULDM. In addition we study the evolution of the velocity field of a solitonic core orbiting within a Newtonian central potential, showing that the stable orbital configuration is an irrotational Riemann-S ellipsoid. Finally, we demonstrate that \PyUltraLight delivers sub-percent level energy conservation for a selection of dynamical scenarios.

\subsection{Interference Patterns During Soliton Collisions}\label{sec:interference}

\begin{figure}
  \includegraphics[trim={0 0 0 0.9cm},clip, scale=0.9]{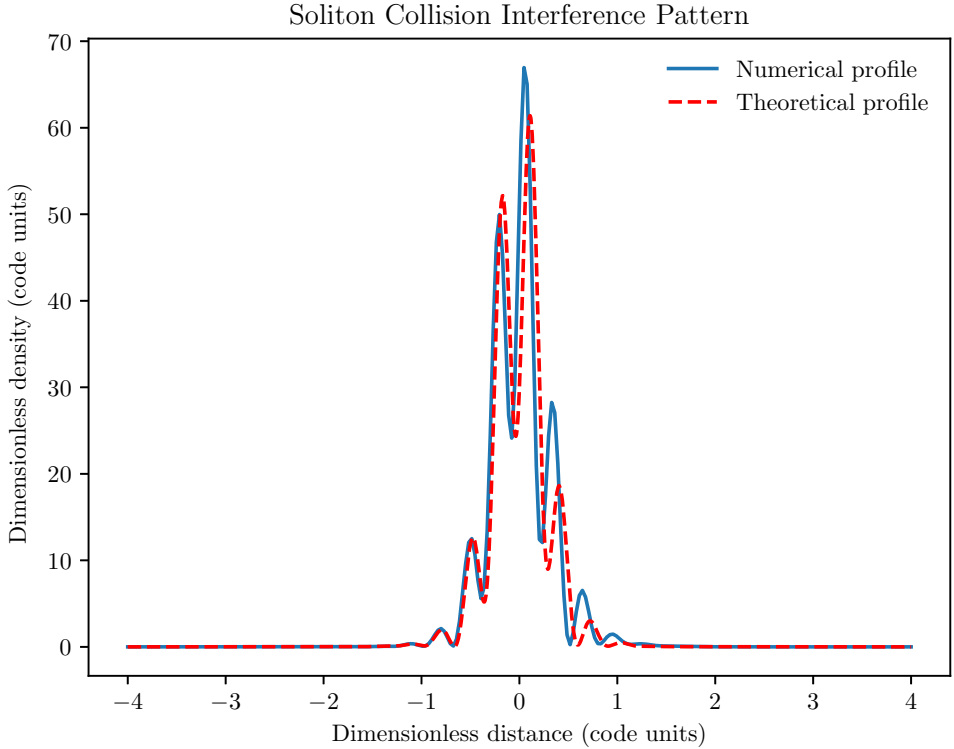}
  \caption{Comparison of theoretical and numerical density profiles at time of maximal interference for head-on collision of two solitons with mass ratio $\mu=2$ and no relative phase difference. The solitons have dimensionless masses 5 and 10, with an initial separation of 4 code units and relative velocity of 20 code units. The simulation resolution is $256^3$ in a box of side-length 8.}
  \label{fig:interference}
\end{figure}

The outcomes of ULDM soliton collisions depend critically on  whether the total energy of the isolated binary system is positive or negative. With a positive total energy the solitons pass through each other, emerging largely undisturbed from their initial configurations and the wavefunctions describing the solitons are superposed during the collision, yielding distinctive interference patterns. 

Following \cite{Schwabe:2016rze}, we consider the head-on collision of two solitions with mass ratio $\mu=2$ and high relative velocity. While we work in dimensionless code units, it should be noted that a dimensionful velocity can be restored from the code velocity by multiplying through by $\CMcal{L}\CMcal{T}^{-1}$, the scale parameters defined in Equations \ref{eq:length} and \ref{eq:time}. This simple case of a head-on soliton collision can be treated approximately. Starting from equation \ref{eq:covariant-solution} we write the total wavefunction of the binary system in terms of dimensionless quantities along the collision axis as
\begin{align}
    \psi(x,t)=& \ \alpha_1 f\big(\sqrt{\alpha_1}\vert x-x_1-v_1 t\vert\big)e^{i\left(\alpha_1\beta t+v_1(x-x_1)-\frac{1}{2}v_1^2 t+\delta\right)}\nonumber\\
    &+\alpha_2 f\big(\sqrt{\alpha_2}\vert x-x_2-v_2 t\vert\big)e^{i\left(\alpha_2\beta t+v_2(x-x_2)-\frac{1}{2}v_2^2 t\right)},
\end{align}
where $x_1$ and $x_2$ are the initial central positions of the solitions, $v_1$ and $v_2$ are the soliton velocities, $\delta$ is a constant relative phase term and $\alpha_2 = 4\, \alpha_1$, parameterising the density profiles as discussed in Section \ref{sec:soliton-profiles}. For convenience we set $v_1=-v_2$ and $x_1=-x_2$. We expect that the interference effects will be maximised when two components of the wavefunction are centred at the same location, such that $x_1+v_1t=-x_2-v_2t=0$. This corresponds to a time $t_{o}=\vert x_1/v_1\vert = \vert x_2/v_2\vert$, where in this simplified model we do not account for distortions caused by the accelerating or compactifying effects that the gravitational interaction has on the soliton profiles as they approach one another. The dimensionless density is then given by
\begin{align}\label{eq:predicted-interference}
    \vert\psi(x,t_{o})\vert^2= \ \alpha_1^2\bigg[&f(\sqrt{\alpha_1}x)^2+16f(2\sqrt{\alpha_1}x)^2+\nonumber\\
    &8f(\sqrt{\alpha_1}x)f(2\sqrt{\alpha_1}x)\operatorname{cos}\left(-3\alpha_1\beta\left\vert\frac{x_1}{v_1}\right\vert+2v_1x+\delta \right)\bigg].
\end{align}
 Figure \ref{fig:interference} shows the dimensionless density profile at the time of maximal interference for two solitons with mass ratio 2 and phase difference $\delta=0$. The numerical result obtained using \PyUltraLight closely matches the theoretical prediction of equation \ref{eq:predicted-interference}. Small disparities between the numerical and theoretical profiles may be attributed to the effect of gravitational contraction not included in the theoretical prediction of equation \ref{eq:predicted-interference} and to a small offset in the true time of maximal interference due to the solitons accelerating as they fall together. We do not expect an exact match, but we have verified that \PyUltraLight qualitatively reproduces the wave interference effects of the ULDM model. With the exception of \cite{Schwabe:2016rze}, few studies of ULDM dynamics have investigated the interference patterns generated by colliding solitons in this way. In some cases, this is because the algorithm employed to simulate the dynamics is not capable of reproducing such effects. An example of this is given in \cite{Veltmaat:2016rxo}, where it is demonstrated that the coarse-grained nature of the particle-mesh method renders the algorithm incapable of reproducing detailed interference patterns such as those shown here. 
 
\subsection{Effective Forces From Destructive Interference}

\begin{figure}
  \includegraphics[width=1.\textwidth, trim={0 0 0 0},clip]{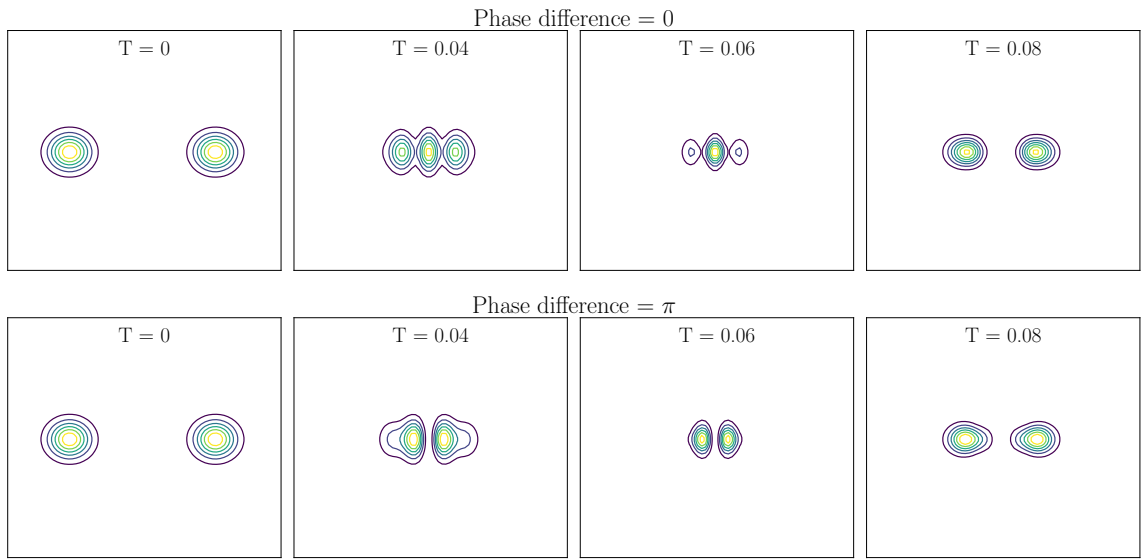}
  \caption{Head-on collisions between solitons of mass 20, initial relative velocity 20, and initial separation 1.2 in code units. Plots show contours of constant density; time progresses from left to right across each row and is indicated in code units for each frame. The upper panel shows solutons initially in phase; the effective repulsive force generated by a $\pi$ phase shift can be seen in the lower panel.}
  \label{fig:repulsion}
\end{figure}

As demonstrated in \cite{Paredes:2015wga}, the wavelike properties of ULDM give rise to effective forces which can dramatically affect the dynamics of core collisions. These effective forces arise as a result of interference phenomena, rather than because of any local interactions the ULDM model might  incorporate.  Figure \ref{fig:repulsion} shows the results of a head-on collision between two solitons, where in one instance the solitons have no initial phase difference, and in the other instance a phase difference of $\pi$ is applied in the initial conditions. In this simulation, solitons of mass 20 are initialised with relative velocity 20 and initial separation 1.2 (code units). The solitons are allowed to collide, and contours of the density profile along the plane of symmetry are displayed. In one case (top) there is no phase offset between the initial solitons, while in the second the phases differ by $\pi$. In the latter case, the phase shift creates an effective repulsive force between the two solitons. It can be seen in the second frame that as the solitons approach one another, the $\pi$ phase shift results in a slowing of the approach accompanied by a deformation of the density profile, acting so as to avoid contact between the solitons. Dissimilarly, in the case where there is no phase shift, the solitons readily collide and merge to form a single contiguous density profile prior to re-separating. Further discussion of this phenomenon and its possible observational consequences can be found in \cite{Paredes:2015wga}.

\vspace{1em}

\subsection{Tidal Disruption of Solitons Orbiting a Central Potential}\label{sec:disruption}

\begin{figure}
  \includegraphics[width=1.\textwidth,trim=1cm 1cm 0 1cm,clip]{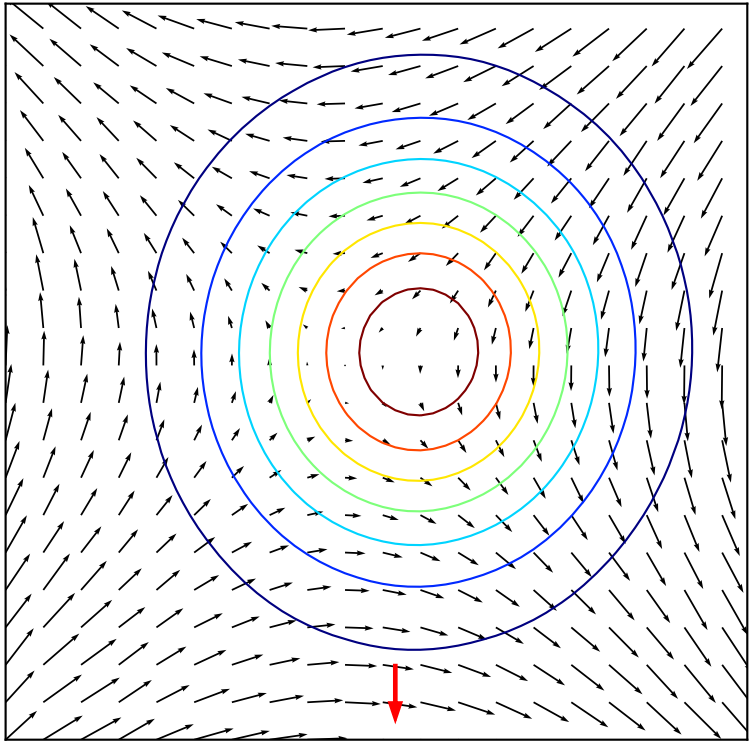}
  \caption{Contours of constant density for a solitonic core after one revolution around a central potential. Contours are superimposed upon the internal velocity field (with the bulk motion subtracted). This velocity field is qualitatively that of an irrotational Riemann-S ellipsoid, and the deformation of the density profile of the soliton along the radial direction (red arrow) is visible. The host:satellite mass ratio is approximately 55; simulation resolution is $256^3$.}
  \label{fig:riemann}
\end{figure}

\PyUltraLight allows the inclusion of a static potential equivalent to a point-mass  at the centre of the simulation region.  There is no backreaction on this mass as a result of the ULDM dynamics, and its ``mirror images'' within the periodic coordinate systems are not accounted for within the overall gravitational potential. While a potential of this form does not necessarily accurately emulate that which we might expect from a realistic galaxy or dark matter halo, it provides a starting point for a study of the stability of satellite dark matter halos orbiting a much larger object. In particular, this includes the investigation of lifespans of dwarf satellite galaxies orbiting much larger objects (including the Milky Way) which are a key to understanding whether ULDM models can resolve the so-called missing satellites problem \cite{Weinberg:2013aya}.

An extensive study of the tidal disruption of ULDM solitonic cores orbiting a central potential has recently been undertaken in \cite{Du:2018zrg} and we reproduce just one of their results here. To do this, we again adopt the definitions \ref{eq:madelung}, namely
\begin{equation}\label{eq:madelung_repeated}
    \psi\equiv\sqrt{\rho}e^{i\theta}, \quad \vec{v}\equiv\boldsymbol{\nabla}\theta,
\end{equation}
where we are working in dimensionless code units. Using these definitions, the Schr{\"o}dinger-Poisson system can be recast in terms of hydrodynamical quantities in the so-called Madelung representation. The Madelung equations resemble the continuity and Euler equations of classical fluid dynamics, with the addition of a `quantum pressure' term accounting for resistance against gravitational collapse. The Madelung formalism is discussed in detail in \cite{Suarez:2011yf, Suarez:2015uva, Johnston:2009wz, Kopp:2017hbb}. Because this hydrodynamical formulation defines the fluid velocity as the gradient of the phase of the field $\psi$, problems arise when $\psi=0$, where the phase is not well defined. Because of this issue, the Madelung and Schr{\"o}dinger representations are not strictly equivalent unless a quantisation condition is imposed, as discussed in \cite{Wallstrom:1994fp}. We do not consider the subtleties of the Madelung representation here, as it is sufficient for our purposes to consider the fluid velocity in the region of a solitonic core, where no field nodes are present.\footnote{It should be noted that, restoring dimensionful units, the fluid velocity $\vec{v}$ is related to the ususal quantum mechanical probability current $\vec{j}$ through $\vert\psi\vert^2\vec{v} = \vec{j} = \hbar/2mi\left[\psi^*\nabla\psi-\psi\nabla\psi^*\right]$} For a discussion of the possible remedies to the `nodes problem', the reader is referred to Chapter 15.3 of \cite{Wyatt:2005uc}.
Where the Madelung representation is well defined, i.e. where the phase is a smoothly varying function, the velocity field of the Schr{\"o}dinger-Poisson system is strictly irrotational, $\nabla\times\vec{v}=0$. If a radially symmetric soliton is initialised in a circular orbit around a Newtonian potential, there will be initial transient behaviour as the spherical profile becomes elongated along the radial direction of the central potential. Meanwhile, the velocity field corresponding to the overall orbital motion of the soliton will be superposed with the internal velocity field, combining so as to produce a net flow with vanishing curl. 

The family of Riemann-S ellipsoids describe non-axisymmetric uniformly rotating bodies whose internal velocity fields have vanishing curl \cite{Chandrasekhar1965}. Therefore, it is the characteristic internal velocity field of a Riemann-S ellipsoid which we expect to arise during our simulation of a soliton orbiting a central mass. It is found in \cite{Du:2018zrg} that an initially spherical solitonic core without self-rotation will gradually spin up to form a tidally-locked ellipsoid with an irrotational internal velocity field when orbiting a host mass. We reproduce this result using \PyUltraLight. Figure \ref{fig:riemann} shows the internal velocity field of a solitonic satellite after one complete revolution around a host mass. The soliton has become elongated along the radial line connecting it to the host, indicating that it is tidally locked, while the velocity field within the tidal radius is visibly irrotational and bears the qualitative trademarks of the Riemann-S ellipsoid as presented in Figure 2 of \cite{RindlerDaller:2011kx}. 

It should be noted that the wider velocity field is not expected to be accurately predicted in a simulation of this kind, though the field within the tidal radius is well-modelled. This is because the initial soliton density profile is defined only out to a given cutoff radius, beyond which the $\psi$ field value is set identically to zero. As mentioned previously, the Madelung hydrodynamical formulation of the Schr{\"o}dinger-Poisson system is not valid where $\psi=0$. Because of this, we focus primarily on the internal velocity field within the high density region of the solitonic core. As we have seen, in this region \PyUltraLight is able to accurately reproduce the expected velocity field characteristics.

\section{Convergence  and Validation}
\subsection{Energy Conservation}\label{sec:energy}

\begin{figure}
  \includegraphics[width=1.1\textwidth,trim=2.5cm 0 0 1cm,clip]{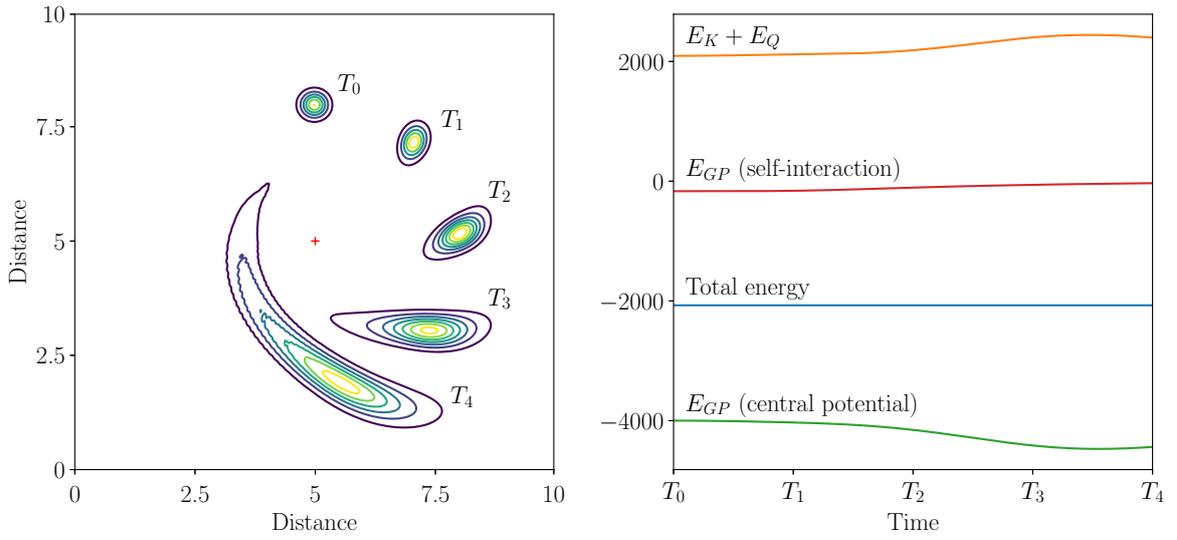}
  \caption{Left: Evolution of the density profile of a solitonic core at equally spaced times as it undergoes tidal disruption in a potential centred at the red cross. Right: Evolution of the components of the total energy of the system; times  correspond to the labeled density profile snapshots. All quantities are in dimensionless code units.}
  \label{fig:combined_1}
\end{figure}

\begin{figure}
  \includegraphics[width=0.9\textwidth,trim=0 0 2cm 2cm,clip]{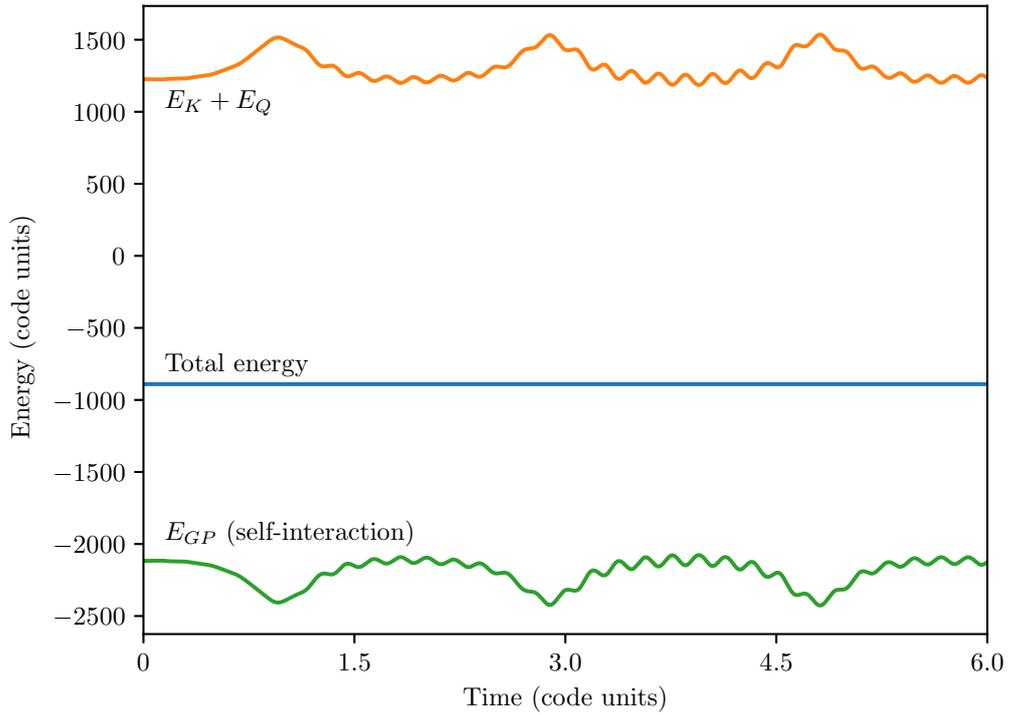}
  \caption{Evolution of the components of the total energy of a binary soliton system with each soliton in an elliptical orbit around the common centre of mass.}
  \label{fig:binary}
\end{figure}

\begin{figure}
  \includegraphics[width=0.9\textwidth,trim=0 0.5cm 0 0,clip]{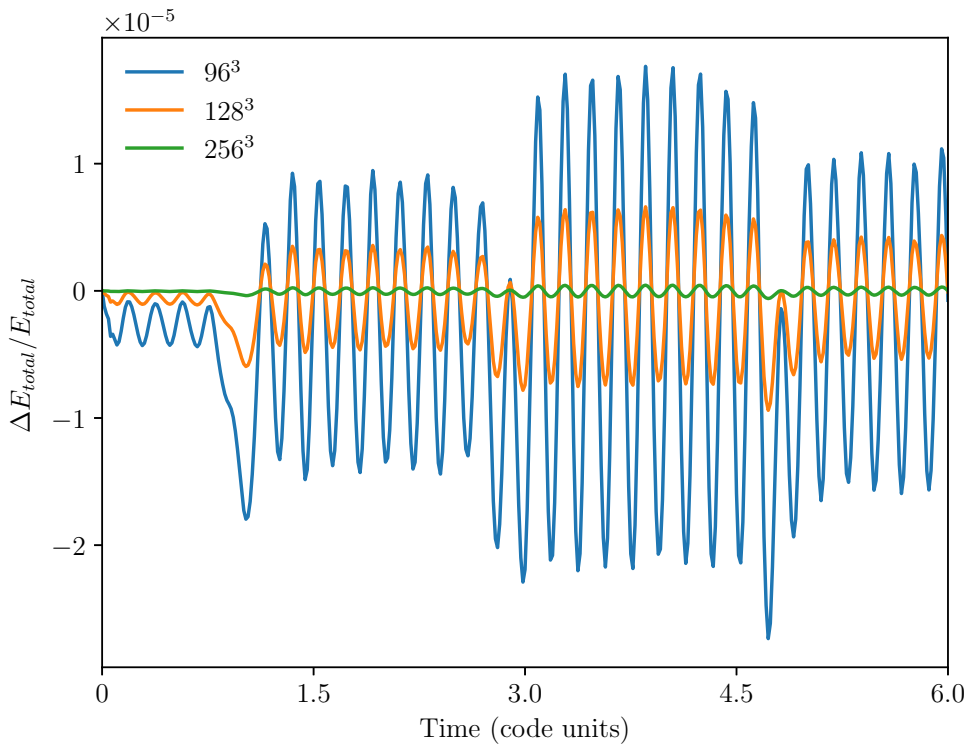}
  \caption{Energy conservation as a function  of  simulation resolution for a binary solitons orbiting their common centre of mass. $\Delta E_{total}$ is the difference between the current total energy and the initial total energy for the configuration. The ratio of this difference to the initial integrated energy is plotted on the y axis for each resolution.}
  \label{fig:energy_change}
\end{figure}

\begin{figure}
  \includegraphics[width=0.9\textwidth,trim=0 0.5cm 0 0,clip]{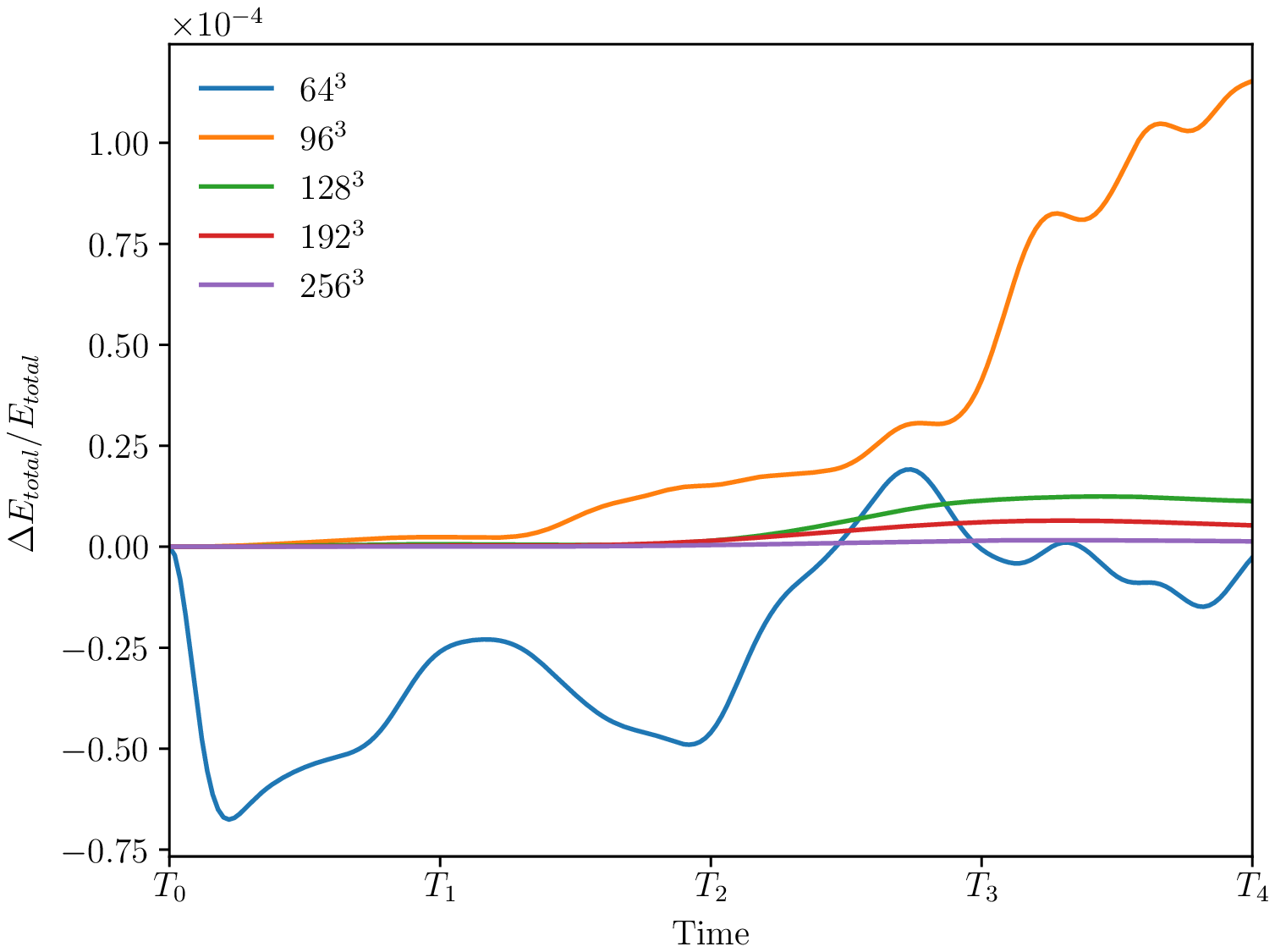}
  \caption{Energy conservation as a function of simulation resolution for a soliton undergoing significant tidal disruption in a Newtonian central potential. $\Delta E_{total}$ is the difference between the current total energy and the initial total energy for the configuration. The ratio of this difference to the initial integrated energy is plotted on the y axis for each resolution.}
  \label{fig:energy_change_2}
\end{figure}

Physically, we expect that the overall energy in the system will be conserved. This provides a test on the numerical performance of \PyUltraLight, and we find that even at relatively low spatial resolution we see sub-percent level energy conservation for all the dynamical scenarios considered here. In this Section we express the energy of the Schr{\"o}dinger-Poisson system in terms of the variables  $\psi$ and $\Phi$ and discuss its decomposition into individual constituents calculated separately within the code. We then present results for a variety of configurations. 

We begin by defining a suitable action which yields the full Schr{\"o}dinger-Poisson system through its corresponding Euler-Lagrange equations. We find that variation of
\begin{equation}\label{eq:sp-action}
    S=\int dt\int_{\mathbb{R}^3} d^3x \  -\bigg\{\frac{1}{2}\vert\nabla\Phi\vert^2+\Phi\vert\psi\vert^2+\frac{1}{2}\vert\nabla\psi\vert^2+\frac{i}{2}(\psi\Dot{\psi}^*-\Dot{\psi}\psi^*)\bigg\}
\end{equation}
with respect to $\Phi$, $\psi^*$ and $\psi$ yields equations \ref{eq:p-adim}, \ref{eq:s-adim}, and the conjugate of equation \ref{eq:s-adim}, respectively. The integrand of equation \ref{eq:sp-action} is the Lagrangian density, $\mathcal{L}$, from which we can derive the conserved energy in the usual way:
\begin{equation}
    E_{tot}=\int_{\mathbb{R}^3}d^3x \ \bigg\{\frac{\partial \mathcal{L}}{\partial \Dot{\psi}}\Dot{\psi}+\frac{\partial \mathcal{L}}{\partial \Dot{\psi}^*}\Dot{\psi}^*+\frac{\partial \mathcal{L}}{\partial \Dot{\Phi}}\Dot{\Phi}-\mathcal{L}\bigg\}.
\end{equation}
Evaluating this expression, we obtain:
\begin{align}
    E_{tot}&=\int_{\mathbb{R}^3}d^3x \ \bigg\{\frac{1}{2}\vert\nabla\Phi\vert^2+\Phi\vert\psi\vert^2+\frac{1}{2}\vert\nabla\psi\vert^2\bigg\}\\
    &=\int_{\mathbb{R}^3}d^3x \ \bigg\{\frac{1}{2}\nabla(\Phi\nabla\Phi)-\frac{1}{2}\Phi\nabla^2\Phi+\Phi\vert\psi\vert^2+\frac{1}{2}\nabla(\psi^*\nabla\psi)-\frac{1}{2}\psi^*\nabla^2\psi\bigg\}\\
    &=\int_{\mathbb{R}^3}d^3x \ \bigg\{\frac{1}{2}\Phi\vert\psi\vert^2-\frac{1}{2}\psi^*\nabla^2\psi\bigg\}.\label{eq:energy-tot}
\end{align}
where in the last step we have used Stokes' Theorem as well as the Poisson equation (\ref{eq:p-adim}) to perform simplifications. Because we are working with the dimensionless quantities defined in equation \ref{eq:dimensionless}, it is easy to see that this quantity is related to the physical energy through multiplication by a constant factor of $\CMcal{L}^5\CMcal{T}^{-4}G^{-1}$. It should be noted that equation \ref{eq:energy-tot} is not equivalent to the expectation value of the Schr{\"o}dinger Hamiltonian, which is itself not a conserved quantity of the Schr{\"o}dinger-Poisson system and is given by
\begin{equation}
    \langle\hat{H}\rangle=\int_{\mathbb{R}^3}d^3x \ \bigg\{\Phi\vert\psi\vert^2-\frac{1}{2}\psi^*\nabla^2\psi\bigg\}.
\end{equation}

The two terms in the integral \ref{eq:energy-tot} are calculated separately within the code. The first term is the gravitational potential energy of the Schr{\"o}dinger-Poisson system, $E_{GP}$. As discussed in \cite{Hui:2016ltb}, the second term may be decomposed into contributions which may be considered separately as kinetic and `quantum' energies, $E_K$ and $E_Q$. However, for our purposes it is sufficient to consider only their combined contribution. When \PyUltraLight includes the central potential of a point mass located at the centre of the simulation grid we have additional energy contributions and calculate the  gravitational potential energy from self-interactions separately from the gravitational potential energy due to the central potential.

Figures \ref{fig:combined_1} and \ref{fig:binary} demonstrate energy conservation for two scenarios. The first case  shows the evolution of the energy of a single soliton undergoing significant tidal disruption within a Newtonian central potential. For this simulation a soliton of mass 12 in code units was initialised at a radial distance of 3 code units from the centre of a Newtonian central potential generated by a central mass of 1000 code units. As the soliton is disrupted, the kinetic energy increases, while the gravitational energy due to the central potential decreases, as expected. Meanwhile, the gravitational potential energy from self-interactions gradually increases toward zero as the disruption continues and the ULDM is halo spread over a greater area. In this case the sum of the individual energy components is conserved to $10^{-4}\ \%$  at a resolution of $256^3$.

Figure \ref{fig:binary} demonstrates the evolution of the energy of a binary system of solitons in elliptical orbits around their common centre of mass over three orbital periods. In dimensionless code units, the soliton masses are 22, the initial separation  2, and the initial relative velocity was 3.6. At points of closest approach the kinetic energy increases as the solitons speed up, while the potential energy due to self-interaction decreases commensurately such that the total energy is conserved. In this scenario no central potential has been included. As the solitons reach the first point of closest approach, they become slightly deformed, exciting oscillatory modes which are manifest in the Figure as small scale oscillations superposed on the global behaviour. Figure \ref{fig:energy_change} demonstrates the relationship between the total integrated energy and the grid resolution for the same binary system of solitons used to generate Figure~\ref{fig:binary}. The vertical axis shows the ratio of the deviation in the total energy to the initial value of the energy, where the deviation is measured as the difference between the current and initial values. Energy is conserved at sub-percent level even at low resolutions ($96^3$), and increasing grid resolution greatly improves accuracy.

Figure \ref{fig:energy_change_2} demonstrates the improvement in energy conservation with increasing grid resolution for a single soliton tidally disrupted in a Newtonian central potential, with the same set up as used in Figure \ref{fig:combined_1}. Namely, a single soliton of mass 12 code units is initialised at a distance of 3 code units from a central mass, $M=1000$. The initial velocity of the soliton is $\sqrt{M/r}$ where $r$ is the radial distance of the soliton from the central mass. The duration of the simulation is 0.5 code units so that the soliton undergoes significant tidal disruption as demonstrated in Figure \ref{fig:combined_1}. While we see that energy is conserved at sub-percent level even for $64^3$ grid resolution, the qualitative behaviour of the mass density distribution in this case is not correct, so we conclude that this resolution is insufficient for convergence despite good energy conservation. This highlights the importance of a multifaceted approach to convergence testing. At $256^3$, energy is conserved to parts in $10^{-6}$.

\subsection{Spatial and Temporal Resolution}\label{sec:resolution}

\begin{figure}
  \includegraphics[width=1.\textwidth,trim=0 0 0 0,clip]{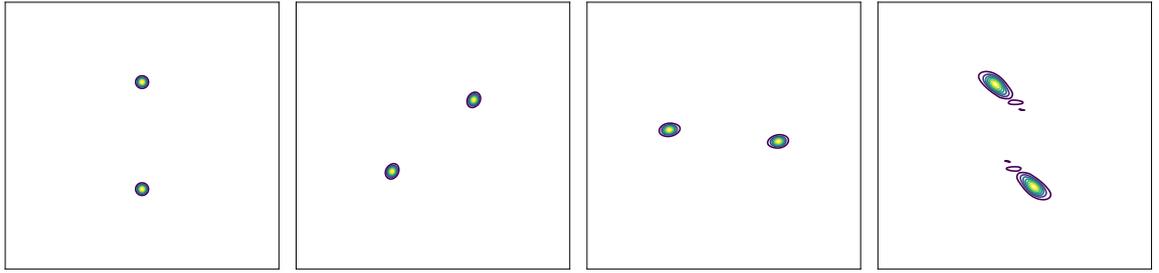}
  \caption{The configuration used to test the sensitivity of solutions to the spatial and temporal resolution. Two solitons of mass $m=20$ are initialised at radial distances $r=2$ from a central mass with $M=1000$ moving in opposite directions with initial speeds $\vert v\vert=\sqrt{M/r}$, corresponding to clockwise orbits around the central mass. The  box size is 10, while the total duration is 0.25 (all quantities in code units). Time runs from left to right. }  \label{fig:disruption}
\end{figure}

We now examine the  convergence of the $\psi$ field configuration as a function of spatial resolution and timestep in a typical simulation. We initialise \PyUltraLight with two  diametrically opposed solitons orbiting a large Newtonian central potential, running until the solitons are tidally disrupted, as shown in Figure \ref{fig:disruption}. 

\begin{figure}
  \includegraphics[width=1.\textwidth,trim=0 1cm 0 2cm,clip]{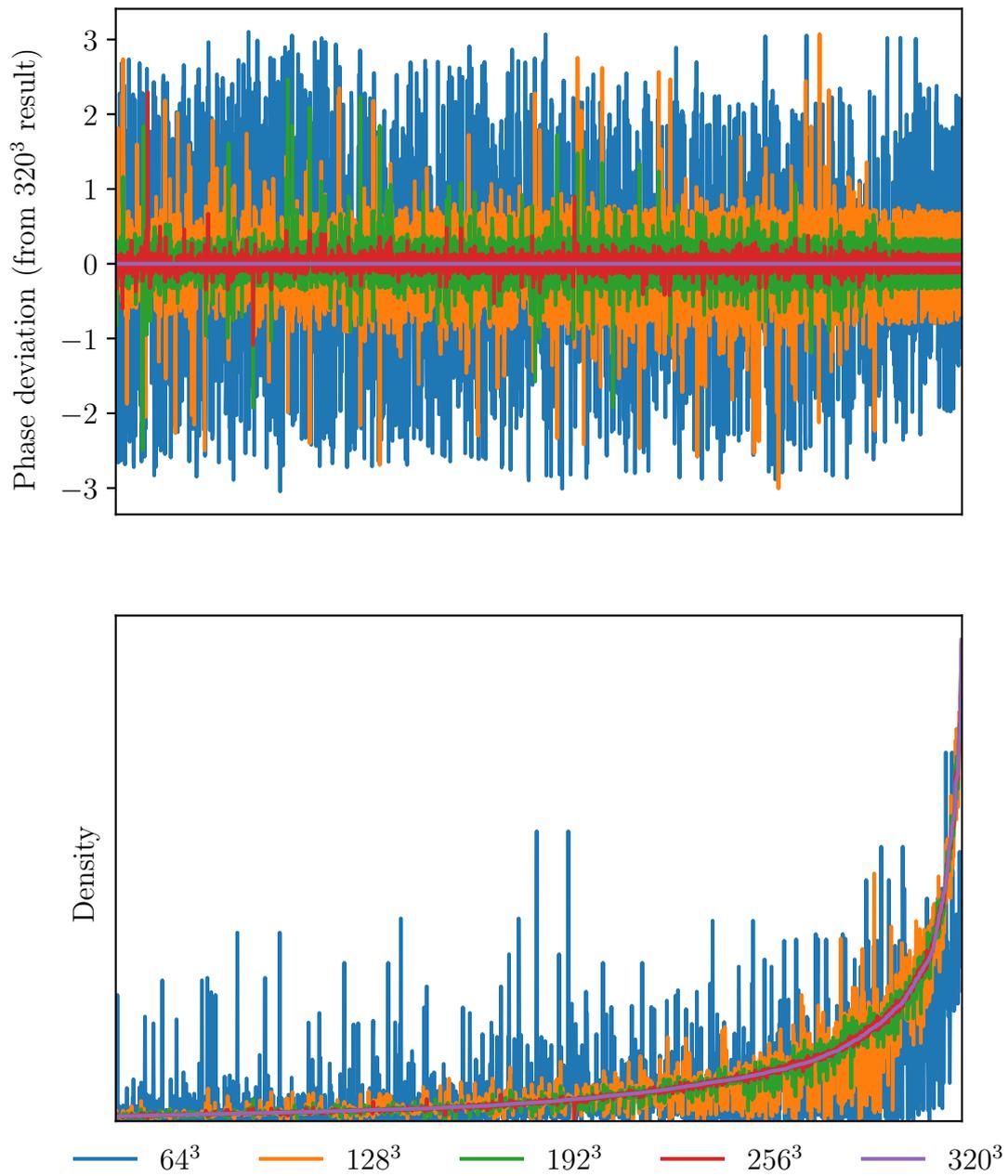}
  \caption{Top: Deviation of the phase of $\psi$ compared to the highest resolution result ($320^3$). Field values are arranged in order of increasing magnitude from left to right. A slight improvement in phase convergence can be seen for higher density regions to the right. Bottom: Improving convergence of $\vert\psi\vert$ with increased spatial resolution for the simulation shown in Figure \ref{fig:disruption}}
  \label{fig:spatial}
\end{figure}

To examine the sensitivity of the  $\psi$ field configuration to the spatial resolution, we first run at $256^3$ with the default timestep. We then re-run  at resolutions  from $64^3$ to $320^3$ with the timestep fixed to the  $256^3$ value and downsample the final outputs to $64^3$. We  sort the resulting values by  the density at the corresponding spatial location, and plot differences in the phase and the magnitude of $\psi$ relative to the values of the $320^3$ run as shown in Figure \ref{fig:spatial} (bottom).  The convergence is poor at $64^3$, but improves with resolution, to the point that there is little difference between the  $256^3$ and $320^3$ cases.

\begin{figure} 
  \includegraphics[width=1.\textwidth,trim=0 1cm 0 2cm,clip]{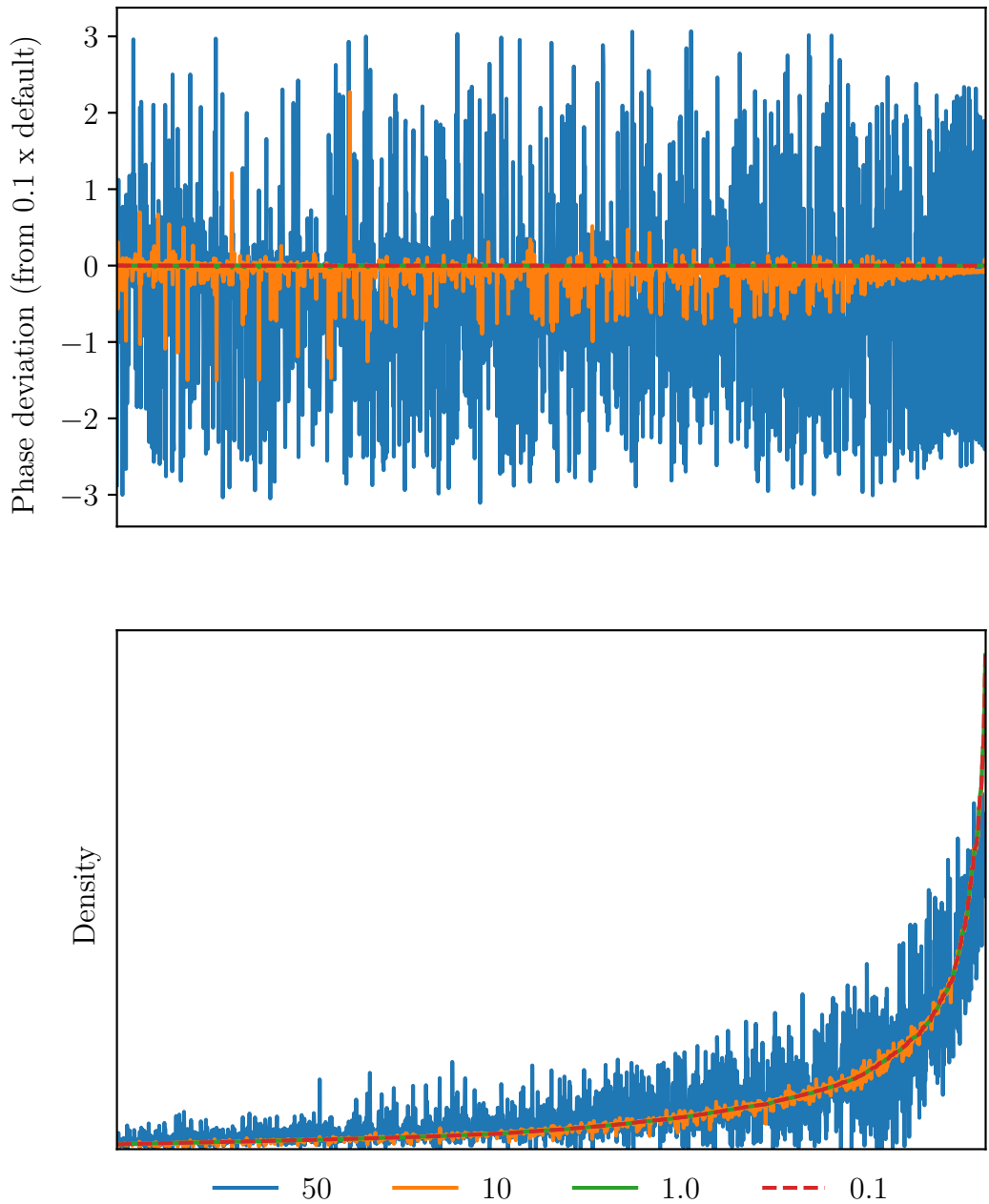}
  \caption{Top: Phase deviation of $\psi$, relative to solution with timestep 0.1 times default, sorted by the density. There is excellent agreement with the default timestep, and reasonable convergence at steps up to $10$ times the default, with better accuracy in high density regions. Bottom: Difference in magnitude of $\psi$, relative to the  solution with timestep 0.1 times default. Again we see good convergence with the default timestep, and tolerable agreement in high density regions when the step is a factor of $10$ or less than default.} 
  \label{fig:temporal}
\end{figure}

To examine the sensitivity of the the  $\psi$ field configuration to the timestep, we take the same default simulation at $256^3$, and then compare this to runs with timesteps 0.1, 10, and 50 times the default and down-sample the final output arrays to $64^3$. We sort sort the array values in order of the $\psi$ field magnitude in the run with the smallest timestep and in Figure \ref{fig:temporal}  we show the difference in the phase and magnitude of $\psi$ as a function of the timestep.  The difference between the results with the default timestep and a value 10 times smaller are negligible; and there is reasonable agreement between the default case and those with the timestep boosted by a factor of 10. However, when the timestep is increased by a factor of $50$ the accuracy of both the phase and magnitude data are significantly reduced.

\begin{figure}
  \includegraphics[width=1.\textwidth,trim=0 1.2cm 0 2.0cm,clip]{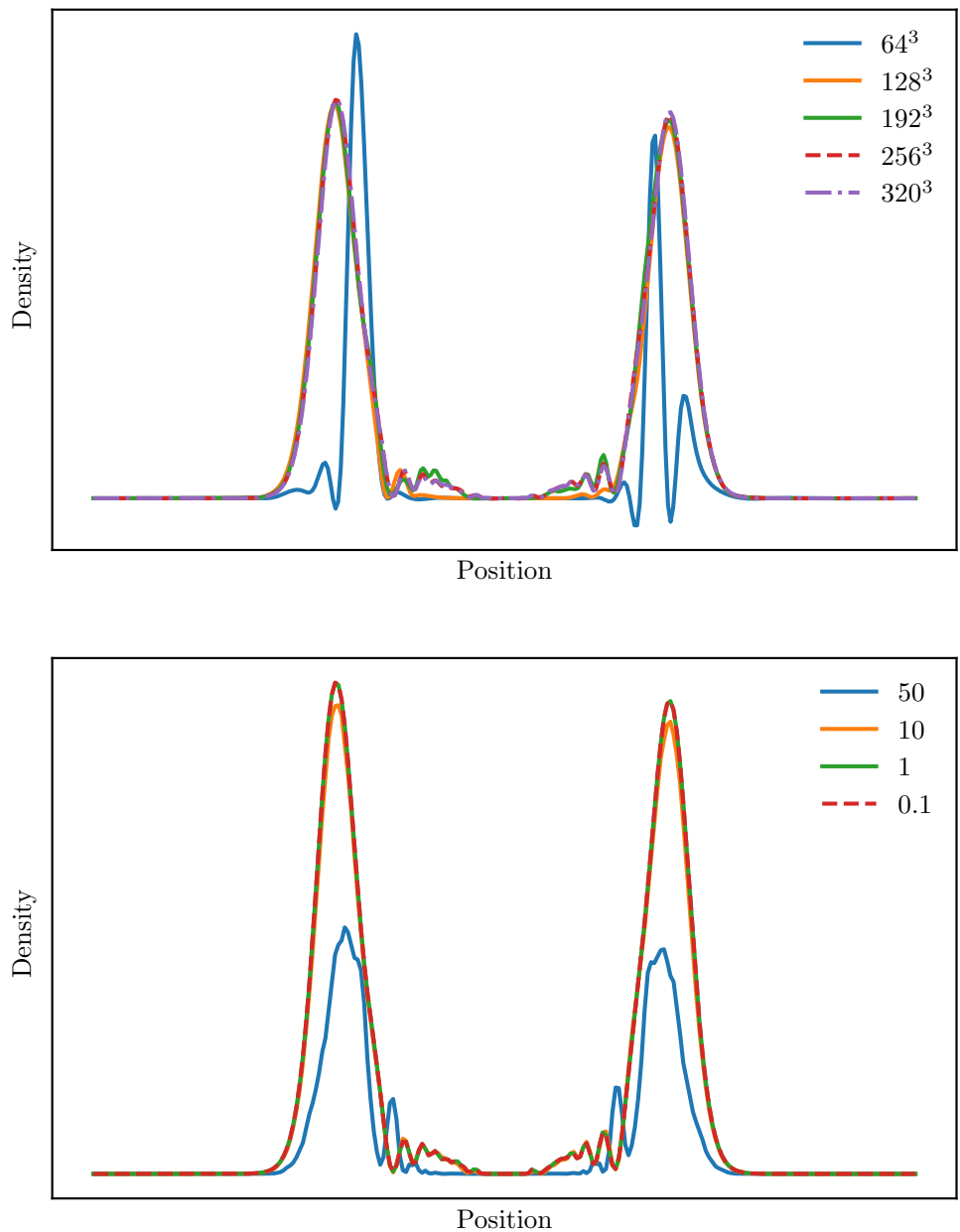}
  \caption{Top: Effect of decreasing the spatial resolution on the density profile at half a revolution. Bottom: Effect of increasing the timestep on the density profile at half a revolution.}
  \label{fig:comparison}
\end{figure}

Figure \ref{fig:comparison} shows profiles of the density through the simulation volume, as a function of spatial resolution and timestep.  Each plot represents the density profile down the axis of symmetry of the initial configuration (vertical axis in Figure \ref{fig:disruption}) after approximately half a revolution around the central potential, or t=0.28 code units -- slightly after the final frame in Figure \ref{fig:disruption} - when the solitons have become distorted due to tidal forces, but are not yet completely disrupted. We see that as the timestep is varied from 0.1 to 50 times the default value, the results with the default and the shorter timestep are virtually indistinguishable, and results are still reasonably accurate at 10 times the default timestep, with small deviations at high densities. However, the results are significantly distorted at 50 times the default timestep. We also see that as the spatial resolution is decreased from $320^3$ to $64^3$, the lowest resolution performs poorly, but there is good convergence  at resolutions of $192^3$ and above. 

\clearpage

\clearpage

\clearpage

\section{Discussion and Outlook}

\PyUltraLight is an accurate, flexible and easy to use tool for  studying the dynamics of ultralight dark matter governed by the Schr{\"o}dinger-Poisson system of equations. The code makes use of a pseudospectral symmetrised split-step Fourier methodology, in which all spatial derivatives are treated via explicit multiplication in the Fourier domain, thereby avoiding difficulties associated with finite-differencing methods. 

Energy conservation within \PyUltraLight is excellent, at sub-percent level for simulations run at $128^3$, with even better performance as resolution is increased. The code captures complex phenomena resulting from the wave-like properties of ultralight dark matter, including the interference patterns arising during high-velocity collisions of solitonic cores, and  effective forces observed in cases where the colliding cores are out of phase. These phenomena can be clearly observed at relatively low spatial resolution, avoiding the need for high-performance computing infrastructure to study the fundamental behaviour of ULDM systems in simple configurations. This makes \PyUltraLight a useful tool for investigating the  dynamics of ULDM systems. 

\PyUltraLight is Python-based, and as such is particularly simple to understand and use. The accompanying Jupyter notebook allows for the efficient adjustment of simulation parameters, and offers a useful browser interface for quick visualisation of simulation results. While Python-based, the code makes use of low-level language resources, namely the FFTW libraries through the use of the Pythonic pyFFTW wrapper and will operate at $\sim 80\%$ efficiency on a 16 core desktop workstation, suggesting that it is computationally efficient. 

The current implementation of \PyUltraLight is already a useful tool for simulating dynamical ULDM systems and exploring their dynamics However, there is much scope for  improvement. In particular, future releases may incorporate a variable timestep and more sophisticated physics, including explicit self-interactions in the axion sector or additional matter components. Augmented versions of the code may also include higher-order generalisations of the pseudo-spectral method, such as those used in \cite{Levkov:2018kau}. \PyUltraLight is publicly available under a BSD license. 

\acknowledgments

We thank Xiaolong Du, Lam Hui, David Marsh, Nathan Musoke, Jens Niemeyer, and Chanda Prescod-Weinstein for valuable discussions on axion / ULDM cosmology, and thank Miro Erkintalo for advice on Schr{\"o}dinger-Poisson solvers  in optical systems.  We acknowledge support from the Marsden Fund of the Royal Society of New Zealand, and  the use of the New Zealand eScience Infrastructure (NeSI) high-performance computing facilities, which are funded jointly by NeSI's collaborator institutions and through the Ministry of Business, Innovation \& Employment's Research Infrastructure programme \url{https://www.nesi.org.nz}.

\appendix
\section{Download and Licensing}

\PyUltraLight is publicly available under a BSD licence. The full repository, including supplementary files such as the code used to generate soliton profiles, is available on GitHub at {\url{https://github.com/auckland-cosmo/PyUltraLight}}. \PyUltraLight makes use of the pyFFTW pythonic wrapper around the FFTW C-based fast Fourier transform libraries. Both pyFFTW and FFTW are freely-available and \PyUltraLight has been used successfully on both Mac OS and Linux systems, as well as a shared-memory cluster environment. We welcome advice and feedback from users.



\bibliographystyle{JHEP-mod}
\bibliography{refs}

\providecommand{\href}[2]{#2}\begingroup\raggedright\begin{thebibliography}{10}

\bibitem{Ade:2015xua}
{\bf Planck} Collaboration, P.~A.~R. Ade et~al., {\it {Planck 2015 results.
  XIII. Cosmological parameters}},  {\em Astron. Astrophys.} {\bf 594} (2016)
  A13, [\href{http://xxx.lanl.gov/abs/1502.01589}{{\tt arXiv:1502.01589}}].

\bibitem{Tan:2016zwf}
{\bf PandaX-II} Collaboration, A.~Tan et~al., {\it {Dark Matter Results from
  First 98.7 Days of Data from the PandaX-II Experiment}},  {\em Phys. Rev.
  Lett.} {\bf 117} (2016), no.~12 121303,
  [\href{http://xxx.lanl.gov/abs/1607.07400}{{\tt arXiv:1607.07400}}].

\bibitem{Akerib:2016vxi}
{\bf LUX} Collaboration, D.~S. Akerib et~al., {\it {Results from a search for
  dark matter in the complete LUX exposure}},  {\em Phys. Rev. Lett.} {\bf 118}
  (2017), no.~2 021303, [\href{http://xxx.lanl.gov/abs/1608.07648}{{\tt
  arXiv:1608.07648}}].

\bibitem{Bull:2015stt}
P.~Bull et~al., {\it {Beyond $\Lambda$CDM: Problems, solutions, and the road
  ahead}},  {\em Phys. Dark Univ.} {\bf 12} (2016) 56--99,
  [\href{http://xxx.lanl.gov/abs/1512.05356}{{\tt arXiv:1512.05356}}].

\bibitem{Hui:2016ltb}
L.~Hui, J.~P. Ostriker, S.~Tremaine, and E.~Witten, {\it {Ultralight scalars as
  cosmological dark matter}},  {\em Phys. Rev.} {\bf D95} (2017), no.~4 043541,
  [\href{http://xxx.lanl.gov/abs/1610.08297}{{\tt arXiv:1610.08297}}].

\bibitem{Kim:2008hd}
J.~E. Kim and G.~Carosi, {\it {Axions and the Strong CP Problem}},  {\em Rev.
  Mod. Phys.} {\bf 82} (2010) 557--602,
  [\href{http://xxx.lanl.gov/abs/0807.3125}{{\tt arXiv:0807.3125}}].

\bibitem{Marsh:2015xka}
D.~J.~E. Marsh, {\it {Axion Cosmology}},  {\em Phys. Rept.} {\bf 643} (2016)
  1--79, [\href{http://xxx.lanl.gov/abs/1510.07633}{{\tt arXiv:1510.07633}}].

\bibitem{Hu:2000ke}
W.~Hu, R.~Barkana, and A.~Gruzinov, {\it {Cold and fuzzy dark matter}},  {\em
  Phys. Rev. Lett.} {\bf 85} (2000) 1158--1161,
  [\href{http://xxx.lanl.gov/abs/astro-ph/0003365}{{\tt astro-ph/0003365}}].

\bibitem{Marsh:2015wka}
D.~J.~E. Marsh and A.-R. Pop, {\it {Axion dark matter, solitons and the
  cusp-core problem}},  {\em Mon. Not. Roy. Astron. Soc.} {\bf 451} (2015),
  no.~3 2479--2492, [\href{http://xxx.lanl.gov/abs/1502.03456}{{\tt
  arXiv:1502.03456}}].

\bibitem{Schwabe:2016rze}
B.~Schwabe, J.~C. Niemeyer, and J.~F. Engels, {\it {Simulations of solitonic
  core mergers in ultralight axion dark matter cosmologies}},  {\em Phys. Rev.}
  {\bf D94} (2016), no.~4 043513,
  [\href{http://xxx.lanl.gov/abs/1606.05151}{{\tt arXiv:1606.05151}}].

\bibitem{Madelung1926}
E.~Madelung, {\it Eine anschauliche Deutung der Gleichung von Schr{\"o}dinger},
   {\em Naturwissenschaften} {\bf 14} (Nov, 1926) 1004--1004.

\bibitem{Zhang:2016uiy}
J.~Zhang, Y.-L.~S. Tsai, J.-L. Kuo, K.~Cheung, and M.-C. Chu, {\it {Ultralight
  Axion Dark Matter and Its Impact on Dark Halo Structure in $N$-body
  Simulations}},  {\em Astrophys. J.} {\bf 853} (2018), no.~1 51,
  [\href{http://xxx.lanl.gov/abs/1611.00892}{{\tt arXiv:1611.00892}}].

\bibitem{Springel:2005mi}
V.~Springel, {\it {The Cosmological simulation code GADGET-2}},  {\em Mon. Not.
  Roy. Astron. Soc.} {\bf 364} (2005) 1105--1134,
  [\href{http://xxx.lanl.gov/abs/astro-ph/0505010}{{\tt astro-ph/0505010}}].

\bibitem{axion-gadget}
``https://github.com/liambx/Axion-Gadget.''

\bibitem{Nori:2018hud}
M.~Nori and M.~Baldi, {\it {AX-GADGET: a new code for cosmological simulations
  of Fuzzy Dark Matter and Axion models}},  {\em Mon. Not. Roy. Astron. Soc.}
  {\bf 478} (2018) 3935, [\href{http://xxx.lanl.gov/abs/1801.08144}{{\tt
  arXiv:1801.08144}}].

\bibitem{Mocz:2015sda}
P.~Mocz and S.~Succi, {\it {Numerical solution of the nonlinear Schr{\"o}dinger
  equation using smoothed-particle hydrodynamics}},  {\em Phys. Rev.} {\bf E91}
  (2015), no.~5 053304, [\href{http://xxx.lanl.gov/abs/1503.03869}{{\tt
  arXiv:1503.03869}}].

\bibitem{Veltmaat:2016rxo}
J.~Veltmaat and J.~C. Niemeyer, {\it {Cosmological particle-in-cell simulations
  with ultralight axion dark matter}},  {\em Phys. Rev.} {\bf D94} (2016),
  no.~12 123523, [\href{http://xxx.lanl.gov/abs/1608.00802}{{\tt
  arXiv:1608.00802}}].

\bibitem{Almgren:2013sz}
A.~Almgren, J.~Bell, M.~Lijewski, Z.~Lukic, and E.~Van~Andel, {\it {Nyx: A
  Massively Parallel AMR Code for Computational Cosmology}},  {\em Astrophys.
  J.} {\bf 765} (2013) 39, [\href{http://xxx.lanl.gov/abs/1301.4498}{{\tt
  arXiv:1301.4498}}].

\bibitem{Benson:2010kx}
A.~J. Benson, {\it {Galacticus: A Semi-Analytic Model of Galaxy Formation}},
  {\em New Astron.} {\bf 17} (2012) 175--197,
  [\href{http://xxx.lanl.gov/abs/1008.1786}{{\tt arXiv:1008.1786}}].

\bibitem{Du:2016zcv}
X.~Du, C.~Behrens, and J.~C. Niemeyer, {\it {Substructure of fuzzy dark matter
  haloes}},  {\em Mon. Not. Roy. Astron. Soc.} {\bf 465} (2017), no.~1
  941--951, [\href{http://xxx.lanl.gov/abs/1608.02575}{{\tt
  arXiv:1608.02575}}].

\bibitem{Springel:2009aa}
V.~Springel, {\it {E pur si muove: Galiliean-invariant cosmological
  hydrodynamical simulations on a moving mesh}},  {\em Mon. Not. Roy. Astron.
  Soc.} {\bf 401} (2010) 791, [\href{http://xxx.lanl.gov/abs/0901.4107}{{\tt
  arXiv:0901.4107}}].

\bibitem{Mocz:2017wlg}
P.~Mocz, M.~Vogelsberger, V.~H. Robles, et~al., {\it {Galaxy formation with
  BECDM - I. Turbulence and relaxation of idealized haloes}},  {\em Mon. Not.
  Roy. Astron. Soc.} {\bf 471} (2017), no.~4 4559--4570,
  [\href{http://xxx.lanl.gov/abs/1705.05845}{{\tt arXiv:1705.05845}}].

\bibitem{Schive:2009hw}
H.-Y. Schive, Y.-C. Tsai, and T.~Chiueh, {\it {GAMER: a GPU-Accelerated
  Adaptive Mesh Refinement Code for Astrophysics}},  {\em Astrophys. J. Suppl.}
  {\bf 186} (2010) 457--484, [\href{http://xxx.lanl.gov/abs/0907.3390}{{\tt
  arXiv:0907.3390}}].

\bibitem{gamer}
``https://github.com/gamer-project/gamer/wiki.''

\bibitem{Schive:2014dra}
H.-Y. Schive, T.~Chiueh, and T.~Broadhurst, {\it {Cosmic Structure as the
  Quantum Interference of a Coherent Dark Wave}},  {\em Nature Phys.} {\bf 10}
  (2014) 496--499, [\href{http://xxx.lanl.gov/abs/1406.6586}{{\tt
  arXiv:1406.6586}}].

\bibitem{Paredes:2015wga}
A.~Paredes and H.~Michinel, {\it {Interference of Dark Matter Solitons and
  Galactic Offsets}},  {\em Phys. Dark Univ.} {\bf 12} (2016) 50--55,
  [\href{http://xxx.lanl.gov/abs/1512.05121}{{\tt arXiv:1512.05121}}].

\bibitem{Easther:2010qz}
R.~Easther, H.~Finkel, and N.~Roth, {\it {PSpectRe: A Pseudo-Spectral Code for
  (P)reheating}},  {\em JCAP} {\bf 1010} (2010) 025,
  [\href{http://xxx.lanl.gov/abs/1005.1921}{{\tt arXiv:1005.1921}}].

\bibitem{Amin:2010dc}
M.~A. Amin, R.~Easther, and H.~Finkel, {\it {Inflaton Fragmentation and
  Oscillon Formation in Three Dimensions}},  {\em JCAP} {\bf 1012} (2010) 001,
  [\href{http://xxx.lanl.gov/abs/1009.2505}{{\tt arXiv:1009.2505}}].

\bibitem{Amin:2011hj}
M.~A. Amin, R.~Easther, H.~Finkel, R.~Flauger, and M.~P. Hertzberg, {\it
  {Oscillons After Inflation}},  {\em Phys. Rev. Lett.} {\bf 108} (2012)
  241302, [\href{http://xxx.lanl.gov/abs/1106.3335}{{\tt arXiv:1106.3335}}].

\bibitem{Zhou:2013tsa}
S.-Y. Zhou, E.~J. Copeland, R.~Easther, et~al., {\it {Gravitational Waves from
  Oscillon Preheating}},  {\em JHEP} {\bf 10} (2013) 026,
  [\href{http://xxx.lanl.gov/abs/1304.6094}{{\tt arXiv:1304.6094}}].

\bibitem{pyfftw}
``https://hgomersall.github.io/pyFFTW/.''

\bibitem{fftw}
M.~Frigo and S.~G. Johnson, ``http://www.fftw.org/.''
\newblock "http://www.fftw.org/".

\bibitem{Young2015}
S.~A.~Q. Young, {\em Signature Changing Spacetimes and WKB Approximations in
  General Relativity}.
\newblock Honors thesis, The University of Texas at Austin, 2015.

\bibitem{Dabo:2008}
I.~Dabo, B.~Kozinsky, N.~E. Singh-Miller, and N.~Marzari, {\it Electrostatics
  in periodic boundary conditions and real-space corrections},  {\em Phys. Rev.
  B} {\bf 77} (Mar, 2008) 115139.

\bibitem{Schive:2014hza}
H.-Y. Schive, M.-H. Liao, T.-P. Woo, et~al., {\it {Understanding the Core-Halo
  Relation of Quantum Wave Dark Matter from 3D Simulations}},  {\em Phys. Rev.
  Lett.} {\bf 113} (2014), no.~26 261302,
  [\href{http://xxx.lanl.gov/abs/1407.7762}{{\tt arXiv:1407.7762}}].

\bibitem{Agrawal2013}
G.~Agrawal, {\em Nonlinear Fiber Optics}.
\newblock Optics and Photonics Series. Academic Press, 2013.

\bibitem{Zhang:2008}
Q.~Zhang and M.~I. Hayee, {\it Symmetrized Split-Step Fourier Scheme to Control
  Global Simulation Accuracy in Fiber-Optic Communication Systems},  {\em
  Journal of Lightwave Technology} {\bf 26} (Jan, 2008) 302--316.

\bibitem{Sinkin2003}
O.~V. Sinkin, R.~Holzlohner, J.~Zweck, and C.~R. Menyuk, {\it Optimization of
  the split-step Fourier method in modeling optical-fiber communications
  systems},  {\em Journal of Lightwave Technology} {\bf 21} (Jan, 2003) 61--68.

\bibitem{Frigo2005}
M.~Frigo and S.~G. Johnson, {\it The Design and Implementation of {FFTW3}},
  {\em Proceedings of the IEEE} {\bf 93} (2005), no.~2 216--231. Special issue
  on ``Program Generation, Optimization, and Platform Adaptation''.

\bibitem{numexpr}
``http://numexpr.readthedocs.io/.''

\bibitem{Ajaib:2013oua}
M.~A. Ajaib, {\it {Numerical Methods and Causality in Physics}},
  \href{http://xxx.lanl.gov/abs/1302.5601}{{\tt arXiv:1302.5601}}.

\bibitem{Taha:1984jz}
T.~R. Taha and M.~J. Ablowitz, {\it {Analytical and Numerical Aspects of
  Certain Nonlinear Evolution Equations. II. Numerical, Nonlinear Schrodinger
  Equation}},  {\em J. Comput. Phys.} {\bf 55} (1984) 203--230.

\bibitem{Suarez:2011yf}
A.~Su{\'a}rez and T.~Matos, {\it {Structure Formation with Scalar Field Dark
  Matter: The Fluid Approach}},  {\em Mon. Not. Roy. Astron. Soc.} {\bf 416}
  (2011) 87, [\href{http://xxx.lanl.gov/abs/1101.4039}{{\tt arXiv:1101.4039}}].

\bibitem{Weinberg:2013aya}
D.~H. Weinberg, J.~S. Bullock, F.~Governato, R.~Kuzio~de Naray, and A.~H.~G.
  Peter, {\it {Cold dark matter: controversies on small scales}},  {\em Proc.
  Nat. Acad. Sci.} {\bf 112} (2015) 12249--12255,
  [\href{http://xxx.lanl.gov/abs/1306.0913}{{\tt arXiv:1306.0913}}].

\bibitem{Du:2018zrg}
X.~Du, B.~Schwabe, J.~C. Niemeyer, and D.~B{\"u}rger, {\it {Tidal disruption of
  fuzzy dark matter subhalo cores}},  {\em Phys. Rev.} {\bf D97} (2018), no.~6
  063507, [\href{http://xxx.lanl.gov/abs/1801.04864}{{\tt arXiv:1801.04864}}].

\bibitem{Suarez:2015uva}
A.~Su{\'a}rez and P.-H. Chavanis, {\it {Hydrodynamic representation of the
  Klein-Gordon-Einstein equations in the weak field limit}},  {\em J. Phys.
  Conf. Ser.} {\bf 654} (2015), no.~1 012008,
  [\href{http://xxx.lanl.gov/abs/1504.01164}{{\tt arXiv:1504.01164}}].

\bibitem{Johnston:2009wz}
R.~Johnston, A.~N. Lasenby, and M.~P. Hobson, {\it {Cosmological fluid dynamics
  in the Schr{\"o}dinger formalism}},
  \href{http://xxx.lanl.gov/abs/0904.0611}{{\tt arXiv:0904.0611}}.

\bibitem{Kopp:2017hbb}
M.~Kopp, K.~Vattis, and C.~Skordis, {\it {Solving the Vlasov equation in two
  spatial dimensions with the Schr{\"o}dinger method}},  {\em Phys. Rev.} {\bf
  D96} (2017), no.~12 123532, [\href{http://xxx.lanl.gov/abs/1711.00140}{{\tt
  arXiv:1711.00140}}].

\bibitem{Wallstrom:1994fp}
T.~C. Wallstrom, {\it {Inequivalence between the Schrodinger equation and the
  Madelung hydrodynamic equations}},  {\em Phys. Rev.} {\bf A49} (1994)
  1613--1617.

\bibitem{Wyatt:2005uc}
R.~E. Wyatt, {\em {Quantum Dynamics with Trajectories. Introduction to quantum
  hydrodynamics}}, vol.~28.
\newblock 2005.

\bibitem{Chandrasekhar1965}
S.~{Chandrasekhar}, {\it {The Equilibrum and the Stability of the Riemann
  Ellipsoids. I.}},  {\em Astrophys. J.} {\bf 142} (Oct., 1965) 890.

\bibitem{RindlerDaller:2011kx}
T.~Rindler-Daller and P.~R. Shapiro, {\it {Angular Momentum and Vortex
  Formation in Bose-Einstein-Condensed Cold Dark Matter Haloes}},  {\em Mon.
  Not. Roy. Astron. Soc.} {\bf 422} (2012) 135--161,
  [\href{http://xxx.lanl.gov/abs/1106.1256}{{\tt arXiv:1106.1256}}].

\bibitem{Levkov:2018kau}
D.~G. Levkov, A.~G. Panin, and I.~I. Tkachev, {\it {Gravitational Bose-Einstein
  condensation in the kinetic regime}},
  \href{http://xxx.lanl.gov/abs/1804.05857}{{\tt arXiv:1804.05857}}.

\end{thebibliography}\endgroup








\end{document}